\documentclass[floatfix,aps,prp,superscriptaddress,twocolumn]{revtex4-2}

\usepackage[T1]{fontenc}
\usepackage{lipsum}  
\usepackage{graphicx}
\usepackage[english]{babel}
\usepackage{upgreek}
\usepackage{amsmath}
\usepackage{amssymb}
\usepackage{tabls}
\usepackage{dsfont}
\usepackage[colorlinks=true,citecolor=blue,linkcolor=red,urlcolor=blue]{hyperref}
\usepackage{bookmark}
\usepackage{physics}

\makeatletter
\def\bbl@set@language#1{%
	\edef\languagename{%
		\ifnum\escapechar=\expandafter`\string#1\@empty
		\else\string#1\@empty\fi}%
	\@ifundefined{babel@language@alias@\languagename}{}{%
		\edef\languagename{\@nameuse{babel@language@alias@\languagename}}%
	}%
	\select@language{\languagename}%
	\expandafter\ifx\csname date\languagename\endcsname\relax\else
	\if@filesw
	\protected@write\@auxout{}{\string\select@language{\languagename}}%
	\bbl@for\bbl@tempa\BabelContentsFiles{%
		\addtocontents{\bbl@tempa}{\xstring\select@language{\languagename}}}%
	\bbl@usehooks{write}{}%
	\fi
	\fi}
\newcommand{\DeclareLanguageAlias}[2]{%
	\global\@namedef{babel@language@alias@#1}{#2}%
}
\makeatother

\newcommand\varpm{\mathbin{\vcenter{\hbox{%
  \oalign{\hfil$\scriptstyle+$\hfil\cr
          \noalign{\kern-.3ex}
          $\scriptscriptstyle({-})$\cr}%
}}}}
\newcommand\varmp{\mathbin{\vcenter{\hbox{%
  \oalign{$\scriptstyle({+})$\cr
          \noalign{\kern-.3ex}
          \hfil$\scriptscriptstyle-$\hfil\cr}%
}}}}

\DeclareLanguageAlias{en}{english}

\begin{document}

\title{Analysis and mitigation of residual exchange coupling in linear spin qubit arrays}

\author{Irina Heinz}
\email{irina.heinz@uni-konstanz.de}
\affiliation{Department of Physics, University of Konstanz, D-78457 Konstanz, Germany}
\author{Adam R. Mills}
\affiliation{Department of Physics, Princeton University, Princeton, New Jersey 08544, USA}
\author{Jason R. Petta}
\affiliation{Department of Physics and Astronomy, University of California, Los Angeles, California 90095, USA}
\affiliation{Center for Quantum Science and Engineering, University of California, Los Angeles, California 90095, USA}
\author{Guido Burkard}
\email{guido.burkard@uni-konstanz.de}
\affiliation{Department of Physics, University of Konstanz, D-78457 Konstanz, Germany}


\begin{abstract}
    In recent advancements of quantum computing utilizing spin qubits, it has been demonstrated that this platform possesses the potential for implementing two-qubit gates with fidelities exceeding 99.5\%. However, as with other qubit platforms, it is not feasible to completely turn qubit couplings off. This study aims to investigate the impact of coherent error matrices in gate set tomography by employing a double quantum dot. We evaluate the infidelity caused by residual exchange between spins and compare various mitigation approaches, including the use of adjusted timing through simple drives, considering different parameter settings in the presence of charge noise.  Furthermore, we extend our analysis to larger arrays of exchange-coupled spin qubits to provide an estimation of the expected fidelity. In particular, we demonstrate the influence of residual exchange on a single-qubit $Y$ gate and the native two-qubit SWAP gate in a linear chain. Our findings emphasize the significance of accounting for residual exchange when scaling up spin qubit devices and highlight the tradeoff between the effects of charge noise and residual exchange in mitigation techniques.

\end{abstract}


\maketitle

\section{Introduction}
Spin qubits \cite{Loss_1998,Burkard2023} have emerged as a promising platform for quantum information processing due to their long coherence times and gate performance \cite{Mills_2022,Xue2022,Noiri2022}. 
Electrostatically formed quantum dots in Si and Ge based heterostructures hosting a single spin of an electron or a hole show potential scalability \cite{PhysRevApplied.18.024053,Tadokoro2021}.
All of the requirements for spin manipulation have been fulfilled, such as electrical control of spin states with intrinsic or extrinsic spin-orbit coupling, and gate voltage control of nearest neighbor exchange coupling. However, only a finite on-off ratio of the exchange interaction is experimentally feasible, since the voltage has to be tuned rather high on short time scales and leaves the issue of residual exchange in qubit devices of several kHz \cite{Takeda_2022, Mills_2022,Noiri_2022_2}. The fidelity of quantum gates is limited in the presence of residual exchange interaction and suffers from correlated errors \cite{Yoneda2023, PhysRevLett.97.207206}. These errors can represent a rather significant issue for quantum error correction \cite{PhysRevA.71.012336,PhysRevA.103.052428} and thus deserve particular attention. 

In our analysis we assume electron spin qubits in silicon, however, for small residual exchange in a strong magnetic field this analysis should be applicable to other spin qubit platforms as well. We consider the effects of residual exchange on idling qubits, as well as single-qubit and two-qubit gates. We calculate gate fidelities and error generators, and then propose ways to mitigate the effects of residual exchange coupling. 
As a starting point we investigate a double quantum dot (DQD) and find optimal driving times and mitigation schemes for the case of two qubits, which can similarly be employed to edge qubits in spin qubit arrays with one-sided residual exchange. We further extend the description to a linear array of spins and expand previous studies on fixed or residual exchange couplings \cite{PhysRevB.101.155301, Kanaar_2021, Kanaar_2022}. We provide a parameter-dependent estimate for the fidelity of single-qubit operations with two residual exchange couplings to nearest neighboring spins.

\begin{figure}
    \centering
    \includegraphics[width=0.48\textwidth]{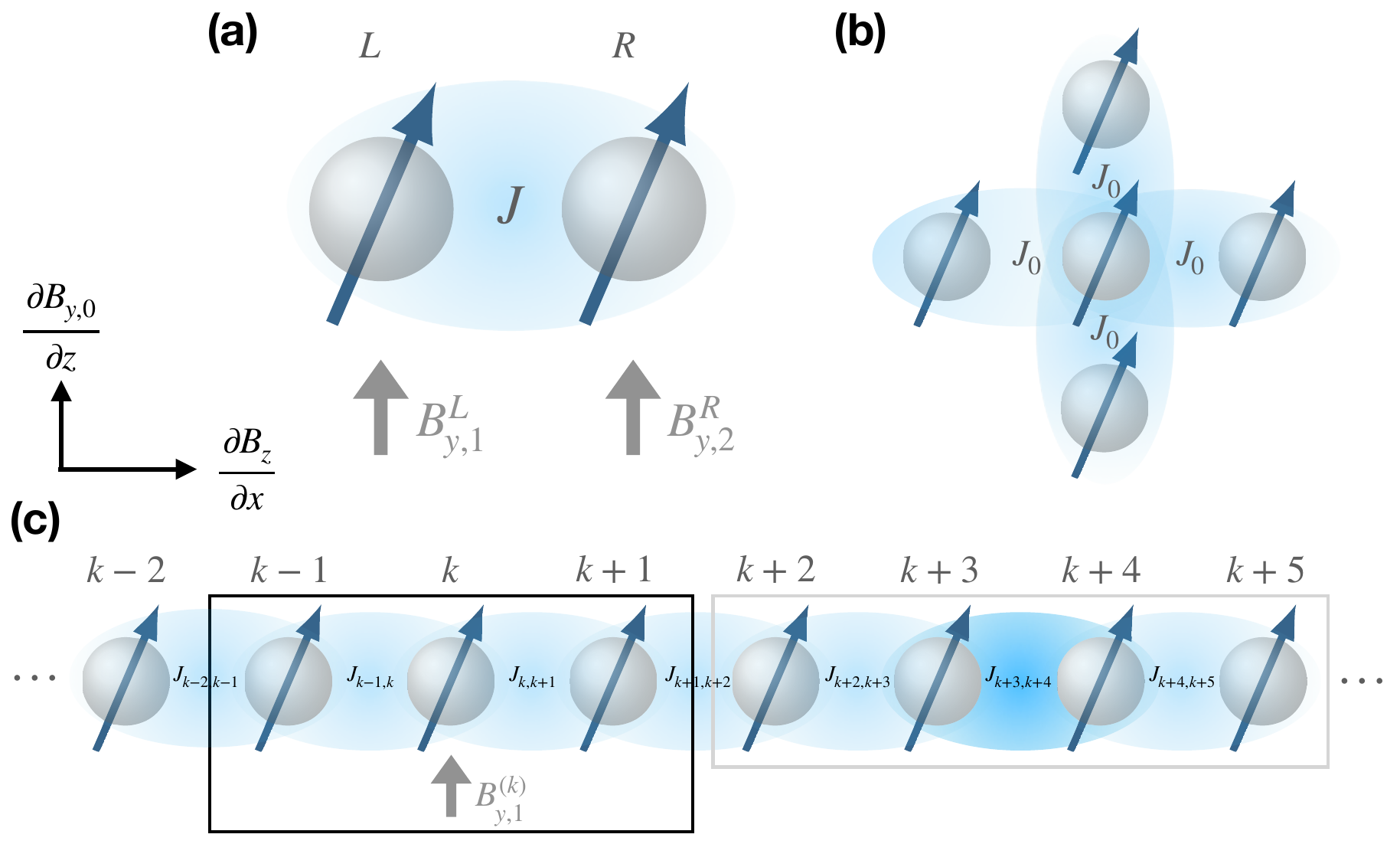}
    \caption{(a) Double quantum dot setup, where $L$ and $R$ label the left and right spin respectively, and $J$ denotes the exchange coupling between the two spins. A gradient of $B_z$ across the array enables individual addressing and the gradient of $B_{y,0}$ along the $z$ direction in the quantum dots enables electrical control of the spin via an effective magnetic driving field $B_{y,1}^L$ and $B_{y,2}^R$. (b) Two-dimensional spin qubit array with nearest neighbor exchange interaction. (c) One-dimensional spin chain with residual exchange when, e.g., driving a single qubit $k$ or performing a CPHASE or a SWAP operation on qubits $k+3$ and $k+4$. The time evolution of a linear qubit array with residual exchange can be separated into the time evolution of the operating qubits and nearest neighbors (black and gray boxes) and $ZZ$ interactions of all other qubit pairs.}
    \label{fig:qubit_array}
\end{figure}

This paper is organized as follows: In Sec.~\ref{sec:DQD} we investigate the DQD case, and discuss mitigation schemes to overcome residual $ZZ$ coupling for a non-driven (Sec.~\ref{Subsec:DQD-nodrive}) and driven (Sec.~\ref{Subsec:DQD+drive}) DQD with one and two drives. 
In Sec.~\ref{Sec:Array} we extend the description to linear spin qubit arrays. We cover the identity, $ZZ$ and $CZ$ operations in Sec.~\ref{Subsec:Array:onlyexchange} and calculate the fidelity of $Y$ gates in Sec.~\ref{Subsec:Array:Ygate}. Finally, we also show the impact of residual exchange in a native SWAP gate in comparison to the impact of the finite magnetic gradient field in Sec.~\ref{Subsec:Array:SWAP}.

\section{Double quantum dot with always-on exchange} \label{sec:DQD}

First we investigate a DQD in the (1,1) charge regime experiencing residual exchange using the two-qubit Hamiltonian
\begin{align}
    H = J (t) \left( \mathbf{S}^L \cdot \mathbf{S}^R -1/4 \right) + \mathbf{S}^L \cdot \mathbf{B}^L + \mathbf{S}^R \cdot \mathbf{B}^R , \label{eq:DQDgeneralHamiltonian}
\end{align}
with the tunable exchange interaction $J(t)$ between spins $\mathbf{S}^{L}$ and $\mathbf{S}^{R}$, and the magnetic field $\mathbf{B}^{\alpha} = (0,B_{y}^{\alpha}(t),B_{z}^{\alpha})$ at the position of spin $\mathbf{S}^{\alpha}$, where $\alpha \in \{L,R\}$ [see Fig.~\ref{fig:qubit_array}(a)]. In addition to the homogeneous magnetic field, a gradient field in the $z$-direction enables individual addressing of the left and right sites of the DQD, $B_{z}^{\alpha}=B_{z} + b_{z}^{\alpha}$. In the $y$-direction, a static gradient field and oscillating plunger gate voltages causes oscillating effective magnetic fields at the quantum dots $B_{y}^{\alpha}(t) = B_{y,0}^{\alpha} + B_{y,1}^{\alpha} \cos(\omega_{1} t + \phi_1) + B_{y,2}^{\alpha} \cos(\omega_{2} t + \phi_2)$. Here $\omega_1 = B_{z}^L$ and $\omega_2 = B_{z}^R$ are the resonance frequencies on the left and right qubits respectively. We define $\Delta B_z = B_{z}^R - B_{z}^L$, $\Delta B_y = B_{y,2}^R - B_{y,1}^L$ and $E_y = B_{y,2}^R + B_{y,1}^L$.

\subsection{Residual exchange without a drive} \label{Subsec:DQD-nodrive}
We first consider the idle evolution, i.e. $B_{y,1}^{\alpha} = B_{y,2}^{\alpha} = 0$. In the rotating frame $\tilde{H}(t) = R^{\dagger}HR+i\dot{R}^{\dagger}R$ with $R=\exp(-i t (B_{z}^L S^L_{z} + B_{z}^R S^R_{z}) )$ we can find a time independent Hamiltonian in the rotating wave approximation (RWA) with $B_{y,0}^{\alpha} \ll B_{z}^{\alpha}$ and $J \ll 2 \Delta B_z$ leading to the time evolution
\begin{align}
    U_0 (t) = & \frac{1}{2} \left(1+e^{i \int \frac{J}{2} \text{d}t} \right) II + \frac{1}{2} \left(1-e^{i\int \frac{J}{2} \text{d}t} \right) ZZ , \label{Eq:U0-2Q}
\end{align}
which equals a CPHASE gate up to a global phase factor and single qubit rotations \cite{Russ2018}. Using the formula $F=(d+|{\rm Tr}[U^\dagger_{\rm ideal}U_{\rm actual}]|^2)/d(d+1)$ for the  fidelity of a unitary gate $U$ with respect to the target gate $U_{\rm ideal}$ in $d=4$ dimensions \cite{Pedersen2007}, we find for the idle operation,
$F_{II}=(1 + |1+\exp(i\int J \text{d}t /2)|^2)/5$. With constant exchange, $J={\rm const.}$, we obtain the idle gate after waiting a time $\tau_{II}=4n\pi/J$ and a $ZZ$ gate after waiting a time $\tau_{ZZ} = 2(2n+1)\pi/J$, of which we make use in the next section.
Here, we work in the rotating frame of the interaction picture with respect to the resonance frequencies of the non-interacting Hamiltonian ($J=0$). We also derive a time evolution in which $J \ll 2 \Delta B_z$ does not hold in Appendix~\ref{App:DQD} explaining GST results with additional Hamiltonian errors, other than $ZZ$ and $II$. However, for always-on residual exchange the change of rotating frames between operations is not trivial and makes timing of pulses important. We stay in the rotating frame of eigenfrequencies of the non-interacting Hamiltonian and restrict ourselves to $J \ll 2 \Delta B_z$ for the remainder of this work, which coincides with the operating regime of most recent experiments.

\subsection{Driven qubits with residual exchange} \label{Subsec:DQD+drive}
\begin{figure*}[t]
	\centering
	\includegraphics[width=0.98\textwidth]{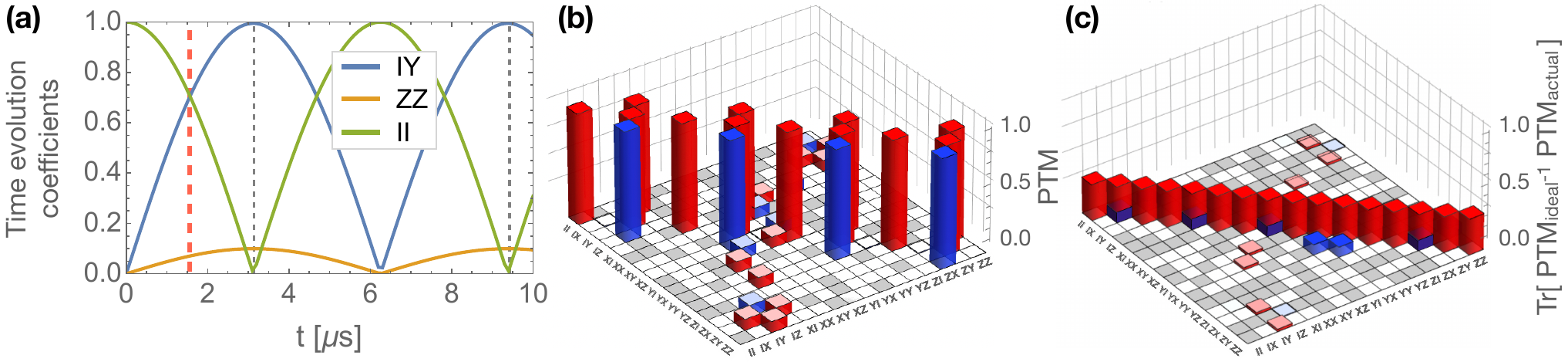}
	\caption{Driven Y-gate on the right qubit with residual exchange. (a) Absolute values of coefficients in Eq.~\eqref{Eq:twodrivesphi00} with $B_{z,1} = 20$~GHz, $B_{z,2} = 20.2$~GHz, $B_{y,1}^L = 0$~MHz, $B_{y,1}^R = 2$~MHz and $J = 200$~kHz. The gray dotted vertical lines correspond to the conventional driving times $\tau = 2 \pi / B_{y,1}^R$ for a $Y$ gate and the red dotted line corresponds to a $\pi/2$ rotation around the $y$ axis. The numerical results perfectly match the analytical approximation in our calculation. (b) Pauli transfer matrix (PTM) and (c) error generator matrix for a $\pi/2$ rotation around the $y$ axis. 
    }
	\label{Fig:YIcoeffs}
\end{figure*}

To perform single qubit gates we resonantly drive both qubits via the terms $ B_{y,1}^{\alpha} \cos(\omega_{1} t + \phi_{1})$ and $B_{y,2}^{\alpha} \cos(\omega_{2} t + \phi_{2})$, with $\omega_{1}=B_{z}^{L}$ and $\omega_{2}=B_{z}^{R}$, as depicted in 
Fig.~\ref{fig:qubit_array}(a). In the rotating frame with $R = \exp(-i t (B_{z}^{L} S^L_{z} + B_{z}^{R} S^{R}_{z}) )$ the  Hamiltonian can be approximated in the RWA in the standard basis $\{ |\uparrow\uparrow~\rangle, |\uparrow\downarrow~\rangle, |\downarrow \uparrow~\rangle, |\downarrow \downarrow~\rangle\}$ as
\begin{widetext}
    \begin{align}
    \tilde{H} = \frac{1}{4} 
    \begin{pmatrix} 
    0 & -i e^{-i \phi_2} B_{y,2}^R & -i e^{-i \phi_1} B_{y,1}^L &0\\ 
    i e^{i \phi_2} B_{y,2}^R & -2J& 0 & -i e^{-i \phi_1} B_{y,1}^L \\
    i e^{i \phi_1} B_{y,1}^L & 0 & -2J & -i e^{-i \phi_2} B_{y,2}^R \\
    0& i e^{i \phi_1} B_{y,1}^L & i e^{i \phi_2} B_{y,2}^R & 0\\
    \end{pmatrix}, \label{eq:doubledot-1Qgate}
\end{align}
\end{widetext}
where we assume $J \ll 2 \Delta B_z$, $B_{y,2}^{\alpha} \ll \Delta B_z$, and $B_{y,1}^{\alpha} \ll 2 B_{z}^{\alpha}$. The corresponding time evolution is given in Eq.~\eqref{Eq:app:twodrivesgeneral} in Appendix~\ref{App:DQD}.
In the case of $Y$ gates, i.e. $\phi_1 = \phi_2 = 0$, we obtain
\begin{equation}\label{Eq:twodrivesphi00}
    \begin{split}
        U (t) =
    &\frac{e^{i\frac{J}{4}t}}{2} \left[
    -i \left(E_y f^+ - \Delta B_y f^- \right)\right. YI  \\
    & 
    -i \left(E_y f^+ + \Delta B_y f^- \right) IY  \\
    & 
    \left.
    +i J  \left(f^+ - f^- \right) XX +\left(g^+ - g^- \right) YY \right. \\
    & 
    \left.
    - i J \left(f^+ + f^- \right)  ZZ 
    + \left(g^+ + g^- \right) II \right],
    \end{split}
\end{equation}
where the full time dependence is contained in the functions
\begin{align}
    f^{\pm} = \frac{1}{\Omega_{y}^{\pm}} \sin\left(\frac{\Omega_y^\pm}{4}t\right) ,\hspace{0.3cm}
    g^{\pm} = \cos\left(\frac{\Omega_y^{\pm}}{4}t\right) \label{eq:fgdef},
\end{align}
and we define $\Omega_{y}^{+} = \sqrt{E_y^2+J^2}$ and $\Omega_y^-= \sqrt{\Delta B_y^2+J^2}$. Fig. \ref{Fig:YIcoeffs} shows the coefficients of Eq.~\eqref{Eq:twodrivesphi00} for a drive, i.e., a $Y$ gate, on the right qubit. 
Analogously, when setting $\phi_1 = \phi_2 = \pi/2$ we obtain 
the time evolution for the $XX$ gate. The time evolution for the gates $YX$ and $XY$ are obtained by setting the respective values for $\phi_1$ and $\phi_2$.

\subsubsection{Drive on a single qubit}
To describe the $IY$ gate we only include one drive $B_{y,2}^{\alpha}$ and set the other drive $B_{y,1}^{\alpha} = 0$. We obtain the time evolution operator in Eq.~\eqref{Eq:app:DQD-1drive} of the Appendix. After the driving time $\tau_{IY} = 2\pi (2n+1)/\Omega_2$, where $\Omega_2=\Omega_{y}^{+}|_{B^{\alpha}_{y,1}=0} 
= \Omega_{y}^{-}|_{B^{\alpha}_{y,1}=0} = \sqrt{(B^{R}_{y,2})^2+J^2}$ and $n \in \mathbb{Z}$, the time evolution results in
\begin{align}
    U_2(\tau_{IY}) = (-1)^{n+1} i e^{i \frac{\pi(2n+1)J}{2\Omega_2}} \left(\frac{B_{y,2}^R}{\Omega_2} IY +\frac{J}{\Omega_2} ZZ  \right) \label{Eq:DQD-IY},
\end{align}
which consists of the desired $IY$ gate and an additional $ZZ$ term. The appearance of the undesired ZZ contribution can be mitigated by choosing $J$ small compared to $B_{y,2}^R$ which remains a hardware optimization problem.

To show the limitations of the fidelity for a single-qubit gate $IX$, $XI$, $IY$ or $YI$ we calculate the upper bound given by $(d + \text{Tr}_{\text{abs}}[U_{\text{actual}} U_{\text{ideal}}^{\dagger}])/(d (d+1))$ as defined in Appendix \ref{App:fidelitybound} and choose $U_{\text{ideal}} = IY$. Since the expression is invariant under diagonal matrix operations one could equally choose $U_{\text{ideal}} = IX$ instead. We find an upper bound for the fidelity of an $IY$ gate to be
\begin{align}
    F_{IY} \leq (1+4 |B_{y,2}^R/\Omega_2|^2)/5 \label{Eq:IYfidbound},
\end{align}
which can be achieved by optimizing the pulse time. Since $|\Delta B_y|\leq |\Omega_y^{-}|$ and $|E_y|\leq |\Omega_y^{+}|$ the limiting fidelity is determined by $J$.

To give an example that can be compared to experimental observations of residual exchange, we calculate the Pauli transfer matrix $\text{PTM} = \text{Tr}[\sigma_j \Lambda(\sigma_i)]$ for a $\pi/2$-rotation around the $y$ axis with $\Lambda(\sigma_i) = U \sigma_i U^{-1} $ and the error generator $\log(\text{PTM}_{\rm ideal}^{-1} \text{PTM}_{\rm actual})$ resulting from solely coherent errors and show the results in Figs.~\ref{Fig:YIcoeffs}(b) and \ref{Fig:YIcoeffs}(c). The driving time for this gate corresponds to the red dotted vertical line in Fig.~\ref{Fig:YIcoeffs}(a). The PTM of an ideal $Y_{\pi/2}$ gate consists of the large bars along the diagonal shown in Fig.~\ref{Fig:YIcoeffs}(b). Here, however, we obtain finite coefficients in the off-diagonals, i.e. at $\{IX, ZY\}$, $\{IY, ZX\}$, $\{IY, ZZ\}$, etc. which are due to the residual exchange. The blue bars indicate negative values while red bars are positive coefficients. The error generator shown in Fig.~\ref{Fig:YIcoeffs}(c) compares the ideal and actual PTM. In experiments, coherent and stochastic errors can be extracted from the error generator using GST \cite{Nielsen_2021, Blume_Kohout_2022}. Here, the error generator only consists of the Hamiltonian projection, i.e. coherent errors, since we do not consider any noise effects.

On the other hand, if a single qubit is driven for a time $2\tau_{IY} = 4\pi n/\Omega_2$ then the time evolution results in a perfect identity gate up to a global phase, $U(2\tau_{IY}) = i (-1)^n \exp(i n \pi J/(\Omega_2)) II$.
This way the $II$ gate does not suffer from an additional $ZZ$ term. Thus when driving one qubit during the waiting time between operations and readout the system is protected from incorporating unwanted $ZZ$ terms. We compare the performance of the drive-induced $II$ gate to the discussed case of no drive in Appendix~\ref{App:DQD-idle} with respect to corrections to the RWA. For large driving fields $B_{y,2}^{R}$,  crosstalk effects from residual driving $B_{y,2}^{L}$ on the left qubit become relevant. Using synchronization \cite{heinz2021crosstalk, PhysRevB.104.045420, PhysRevB.105.L121402} and smooth pulse protocols \cite{Barnes_2015, PhysRevB.101.155301} one can compensate  crosstalk errors during gate operations.

\begin{figure}[t]
    \centering
    \includegraphics[width=0.48\textwidth]{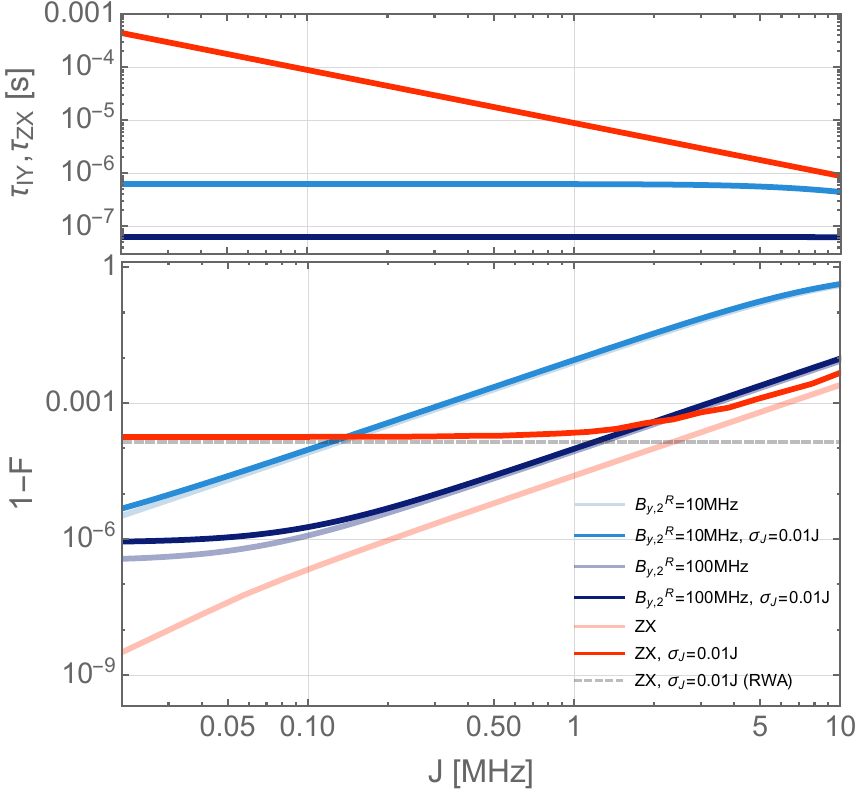}
    \caption{Realizations of a single-qubit Y gate. Driving time (upper plot) and numerical infidelity (lower plot) for the single-drive (blue) and the multi-drive IY gate via the auxiliary ZX (red), as a function of the exchange $J$ in a DQD. Single-drive gate $U_2(\tau_{IY})$ as in Eq.~\eqref{Eq:DQD-IY} with $B_{y,2}^R =$ 10 MHz and 100 MHz (blue) is compared to the mitigation sequence Eq.~\eqref{Eq:DQD-IYIYmain} (red) when quasi-static Gaussian charge noise with standard deviation $\sigma_J = 0.01 J$ is present. Lines in light color show the noise-free behaviour with a slope determined by the validity of the approximations made ($B_{y,2}^{R}, J \ll \Delta B_z$). The horizontal dashed gray line gives the result for the infidelity due to Gaussian noise with $\sigma_J = 0.01 J$ within the RWA. For the numerical calculations the following parameters were used: $B_z^L=20$ GHz, $B_z^R=20.2$ GHz, $B_{y,0}^L=0.1$ MHz, $B_{y,0}^R=0$ MHz.}
    \label{fig:DQDfid}
\end{figure}
\subsubsection{Error mitigation for single-qubit gates} \label{SubSec:DQD-mitigate}
We have shown that one-shot single-qubit gates of type $IX$, $XI$, $IY$ or $YI$ cannot be ideally implemented with residual exchange.
Although a sequence including ideal $z$-rotations could mitigate residual $ZZ$ terms, note that $IZ$ and $ZI$ rotations are obtained by incorporating the respective phase into the drive and thus leaves the residual $ZZ$ term after the sequence (see Appendix~\ref{App:DQD-mitigate-1Q} for more details). 
However, when applying $U_2(\tau_{IY})$ twice in a row (see Eq.~\eqref{Eq:App:IYIYgeneral} for full time evolution) and choosing the residual exchange for the second shot equal to the first one $\tilde{J} = J$, but  driving with a $\pi$-phase shift for the second unitary, such that $\tilde{B}_{y,2}^R = - B_{y,2}^R$, we can write
    \begin{align}
    \tilde{U}_2(\tilde{\tau}_{IY}) U_2(\tau_{IY}) &=
    - \frac{(-1)^{n+\tilde{n}}}{\Omega_2^2} e^{i \frac{\pi(n+ \tilde{n}+1)J}{\Omega_2}} \label{Eq:DQD-IYIYmain}\\
    \times & \left[ \left(- (B_{y,2}^R)^2 + J^2 \right) II + i \left(- 2B_{y,2}^R J \right) ZX \right], \nonumber
\end{align}
with $n, \tilde{n} \in \mathbb{Z}$. If $J=B_{y,2}^R$ the $II$ term disappears and the remaining $ZX$ term can be used to compose $IY=(ZZ)(ZX) = e^{i\varphi} U_0(\tau_{ZZ}) \tilde{U}_2(\tilde{\tau}_{IY}) U_2(\tau_{IY})$ up to a global phase $\varphi$ using $U_0$ as described in Sec.~\ref{Subsec:DQD-nodrive}. The same procedure is possible with different combinations of phases $\phi_1$ and $\phi_2$. To demonstrate the validity for this application we calculate and compare the numerical infidelity of a single drive as in Eq.~\eqref{Eq:DQD-IY} for driving amplitudes $B_{y,2}^R= 10$, $100$ MHz (blue) and the $ZX$ composed as in Eq.~\eqref{Eq:DQD-IYIYmain} (red) for increasing exchange value (with $B_{y,2}^R = J$) in Fig.~\ref{fig:DQDfid}. The upper plot shows the corresponding driving times. The light colored lines represent the numerical result without noise, while the dark colors show the performance in the presence of quasi-static Gaussian charge noise on the exchange value $J$ with standard deviation $\sigma_J = 0.01 J$.The flattening of the dark blue curve towards low exchange values is explained by corrections due to the RWA ($J\ll \Delta B_z$), and the constant offset for the noisy $ZX$ case (gray dashed line) is determined by the ratio $\sigma_J /J$ as discussed in Appendix~\ref{Sec:DQDnoise}. The slope of the blue curves is well described by the analytic expression in Eq.~\eqref{Eq:IYfidbound}. The slope of the red curves, however, arise due to corrections to the RWA used to obtain Eq.~\eqref{eq:doubledot-1Qgate}. We find that despite the longer driving time, the mitigation technique composing a $ZX$ gate performs better than the single drive with low driving amplitude $B_{y,2}^R$. Further calculations indicate that if including fluctuations of the drive amplitude due to charge noise, this advantage is still present.

Note that in gate set tomography (GST) usually $X$ and $Y$ gates refer to a $\pi/2$ rotation, i.e. $IX_{\pi/2} = \exp(i \sigma_x \pi / 4) = (IX + II)/\sqrt{2}$ and $IY_{\pi/2} = \exp(i \sigma_y \pi / 4) = (IY + II)/\sqrt{2}$. To obtain the respective gates one can adapt the values of $J$, $\Delta B_{y}$ and $E_y$ such that a certain part of the identity $II$ remains in the Hamiltonian. As an explicit example we can obtain $IY_{\pi/2}=(ZZ)(ZX+ZZ)/\sqrt{2} = e^{i\varphi} U_0(\tau_{ZZ}) \tilde{U}_2(\tilde{\tau}) U_2(\tau)$ with global phase $\varphi$, where $\tilde{\tau} = \tau = 4(\arccot(B_2/\Omega_2) + n \pi)/\Omega_2$, $n\in \mathbb{Z}$, and $\tilde{J}=J$, $\tilde{B}_y=B_y$, and $J=\sqrt{2} B_y$.

\subsubsection{Drive on both qubits}
To perform the $YY$ gate we consider Eq.~\eqref{Eq:twodrivesphi00}. Here we find cosine functions for the $YY$ and $II$ terms and sine functions for all other contributions. If we require the driving time $\tau_{YY}$ to be $\tau_{YY} = 4\pi n/\Omega_y^+$ and simultaneously $\tau_{YY} = 4\pi m/\Omega_y^-$, i.e. $4\pi n/\Omega_y^+ = 4\pi m/\Omega_y^-$, with integers $n$ and $m$, then we find the condition
\begin{align}
    J=\sqrt{\frac{m^2 E_y^2 - n^2 \Delta B_y^2}{n^2-m^2}}. \label{Eq:JcondYY}
\end{align}
Inserting Eq.~\eqref{Eq:JcondYY} into the time evolution in Eq.~\eqref{Eq:twodrivesphi00} leads to 
\begin{equation} \label{Eq:YY}
    \begin{split}
        U (\tau_{YY}) = \frac{1}{2} e^{i\frac{n\pi J}{\Omega_y^+}}  & \left( \left[(-1)^n - (-1)^m \right] YY \right. \\
        & \hspace{0.cm} \left. + \left[(-1)^n + (-1)^m \right] II \right).
    \end{split}
\end{equation}
For $n\cdot m$ even, i.e. $(-1)^n = (-1)^m$, the time evolution results in the identity operation with a global phase factor and also provides a cancellation of the $ZZ$ terms which arise without a drive. However, for $n\cdot m$ odd, i.e. $(-1)^n = - (-1)^m$ we obtain an ideal $YY$ gate with an overall global phase factor. The same arguments hold for the $XX$, $YX$ and $XY$ gates. The robustness of this gate  in the presence of charge noise is similar to the behaviour of $ZX$ gate in Fig.~\ref{fig:DQDfid}. Hence, the overall slope again determines the validity of the RWA and the amount of noise sets the level to which the fidelity saturates and thus the validity of the protocol compared to the naive implementation.

\section{Linear qubit arrays with residual exchange} \label{Sec:Array}
So far we have only considered the performance of a DQD with residual exchange. However, when scaling up spin qubit devices, a finite exchange between neighboring spins will be present. Therefore, we consider a chain of $N$ qubits in a magnetic gradient field, in which only nearest neighbors $\langle i,j \rangle$ are connected by exchange $J_{ij}$ as in Fig.~\ref{fig:qubit_array}(c). We note that next-nearest neighbor interactions are small compared to the already small residual nearest neighbor interaction \cite{Braakman_2013}. We write the Hamiltonian as
\begin{align}
    H (t) = \sum_{\langle i,j \rangle}^{N} J_{ij} (t) \left( \mathbf{S}^{(i)} \cdot \mathbf{S}^{(j)} -1/4 \right) + \sum_{i=1}^{N} \mathbf{S}^{(i)} \cdot \mathbf{B}^{(i)}, \label{Eq:NqubitH}
\end{align}
with the magnetic field $\mathbf{B}^{(i)} = (0,B_{y}^{(i)}(t),B_{z}^{(i)})$ at the position of spin $\mathbf{S}^{(i)}$, where $i,j \in \{1, \dots, N\}$. We have a homogeneous and gradient field in the $z$-direction $B_{z}^{(i)}=B_{z} + b_{z}^{(i)}$ for individual addressing. In the $y$-direction a static gradient field which together with oscillating gate voltages creates oscillating magnetic fields at the QDs, $B_{y}^{(i)}(t) = B_{y,0}^{(i)} + B_{y,1}^{(i)} \cos(\omega^{(i)} t + \phi^{(i)})$.

\subsection{Residual exchange}\label{Subsec:Array:onlyexchange}
First, we only consider the undriven system, i.e. $B_{y,1}^{(i)} = 0$ for all $i$. Following Sec.~\ref{Subsec:DQD-nodrive}, in the rotating frame $\tilde{H}(t) = R^{\dagger}HR+i\dot{R}^{\dagger}R$ with $R=\exp(-it \sum_{i}^N B_{z}^{(i)} S^{(i)}_{z} )$ we can find a time-independent diagonal Hamiltonian in the RWA $\tilde{H} \approx \sum_{\langle i,j \rangle} J_{ij} (S_z^{(i)} S_z^{(j)} - 1/4)$ if $|J_{ij}| \ll 2 |B^{(i)}_{z} - B^{(j)}_{z}|$ leading to the time evolution
\begin{equation}
    \begin{split}
        U (t) = 
        \prod_{\langle i,j \rangle} \frac{1}{2} & \left( \left[ \exp(i\frac{J_{ij}}{2} t) +1 \right] I \right.\\
        &\hspace{0.2cm} \left. - 4  \left[ \exp(i\frac{J_{ij}}{2} t) -1 \right] S_z^{(i)} S_z^{(j)} \right). \label{Eq:Nonlyexchange}
    \end{split}
\end{equation}
When expanding the product in Eq.~ \eqref{Eq:Nonlyexchange} we find the first term to be the identity and all other terms to contain at least one $S_z^{(i)}$ in the tensor product. Using properties of the tensor product and the trace, the fidelity for the time evolution being the identity gate is then given by
\begin{align}
    F_I 
    = \frac{1 + d\left| \prod_{\langle i,j \rangle} \frac{1}{2} \left[ \exp(i\frac{J_{ij}}{2} t) + 1 \right] \right|^2}{d+1} ,
\end{align}
with Hilbert space dimension $d=2^N$.
Apparently, the residual exchange between qubits leads to a tunable $N$-qubit CPHASE-like gate. For the idle operation on all qubits the exchange couplings can be tuned to the same value $J_{ij} = J$ and after waiting time $t=4\pi n/J$ with integer $n$ the time evolution equals the identity. Otherwise the time evolution constantly changes the relative phases between the states of neighboring qubits or results in $Z$ gates on part of the array. This can be used for two-qubit gates between neighbouring gates or to perform $z$-rotations on some of the qubits.

\subsection{Driven qubit arrays} \label{Subsec:Array:Ygate}
Next, we add the drive to the Hamiltonian Eq.~\eqref{Eq:NqubitH}. In the rotating frame described by $R=\exp(-it \sum_{i}^N \omega^{(i)} S^{(i)}_{z})$, the Hamiltonian is 
\begin{align}\label{Eq:HJ+HB+Hz}
    \tilde{H} \approx H_J + H_B + H_z ,
\end{align}
where $H_B$ describes the driving term
\begin{align}
    H_B = \sum_{i} \frac{B_{y,1}^{(i)}}{2} \left( \cos(\phi^{(i)}) S_{x}^{(i)} - \sin(\phi^{(i)}) S_{y}^{(i)} \right),
\end{align}
with $2|B_{y,0}^{(i)}|,|B_{y,1}^{(i)}| \ll 2 \omega^{(i)}$, and $H_z$ includes the remaining $B_z$ contribution
in the rotating frame,
\begin{align}
    H_z = \sum_{i} \left(B_{z}^{(i)}-\omega^{(i)}\right) S_z^{(i)}.
\end{align}
The first part of Eq.~\eqref{Eq:HJ+HB+Hz} is due to the exchange interaction, which can be approximated in the far off-resonant regime of two neighboring driving frequencies $|\omega^{(i)}-\omega^{(j)}|\gg J_{ij}$ as
\begin{align}
    H_J = \sum_{\langle i,j \rangle} J_{ij} (S_z^{(i)} S_z^{(j)} - 1/4), \label{Eq:HJ1}
\end{align}
and in the resonant regime $\omega^{(i)}=\omega^{(j)}$, i.e. neighboring qubits are driven with the same frequency, as
\begin{align}
    H_J = \sum_{\langle i,j \rangle} J_{ij} \left(S_z^{(i)} S_z^{(j)} + \frac{1}{2} \left( S_{+}^{(i)} S_{-}^{(j)} + S_{-}^{(i)} S_{+}^{(j)} \right) - \frac{1}{4} \right). \label{Eq:HJ2}
\end{align}
A combination of some off-resonantly and other resonantly driven qubit pairs leads to a combination of the expressions in Eqs.~\eqref{Eq:HJ1} and \eqref{Eq:HJ2}. The case in which $\omega^{(i)} = B_z^{(i)}$, which corresponds to driving each qubit with its own resonance frequency, and hence, $|\omega^{(i)}-\omega^{(j)}| = |B_z^{(i)}-B_z^{(j)}| \gg J_{ij}$, the Hamiltonian becomes 
\begin{equation}\label{Eq:TransIsing}
    \begin{split}
        \tilde{H} \approx & \sum_{\langle i,j \rangle} J_{ij} (S_z^{(i)} S_z^{(j)} - 1/4) \\
        & - \sum_{i} \frac{B_{y,1}^{(i)}}{2} \left( \sin(\phi^{(i)}) S_{x}^{(i)} - \cos(\phi^{(i)}) S_{y}^{(i)} \right). 
    \end{split}
\end{equation}

\subsubsection{Single driven qubit} 
If only a single qubit is driven, note that $J_{ij}$ terms containing only non-driven qubits $i$ and $j$ can be separated in the time evolution since in that case $S_z^{(i)}$ and $S_z^{(j)}$ commute with all other operators and lead to a similar expression as in Eq.~\eqref{Eq:Nonlyexchange}. The total time evolution is given by the product of the latter and the nontrivial matrix exponential of the remaining term containing the driven qubits. To show an example we assume a drive on a qubit $1 < k < N$ as depicted in the black box in Fig.~
\ref{fig:qubit_array}(c) (for qubits 1 and $N$ there is only one neighbor, which simplifies the time evolution to the previously discussed case of a DQD) and write the time evolution as $U(t) = U_{i,j\neq k} (t) U_{k} (t) $ with
\begin{widetext}
    \begin{align}
        U_{i,j\neq k} (t) = \prod_{\langle i,j \rangle,  i,j \neq k} \exp(i\frac{J_{ij}}{4} t) \left( \cos(\frac{J_{ij}}{4} t) I - 4 i \sin(\frac{J_{ij}}{4} t) S_z^{(i)} S_z^{(j)} \right), \label{Eq:Uijnoteqk}
    \end{align}
    \begin{equation}
    \begin{split}
        U_k(t) &= e^{i t\frac{J_{k-1, k}+J_{k, k+1}}{4} } \frac{1}{2} \left(
        -i B_{y,1}^{(k)} \left( f_+ + f_- \right) Y^{(k)}  - i B_{y,1}^{(k)} \left( f_+ - f_- \right) Z^{(k-1)}Y^{(k)}Z^{(k+1)}  \right. \\
        &\hspace{3cm} \left.
        -i \left( E_J f_+ - \Delta J f_- \right) Z^{(k-1)} Z^{(k)} 
        -i  \left( E_J f_+ + \Delta J f_- \right) Z^{(k)} Z^{(k+1)} \right. \\
        &\hspace{3cm} \left.
        +i \left( g_+ - g_- \right) Z^{(k-1)} Z^{(k+1)} 
        + \left( g_+ + g_- \right) I
        \right) , \label{Eq:arrayYgate-time-evolution}
    \end{split}
    \end{equation}
\end{widetext}
where $Y^{(i)}$ and $Z^{(i)}$ are the Pauli operators on qubit~$i$. Here we defined $E_J = J_{k-1, k} + J_{k, k+1}$, $\Delta J = J_{k, k+1} - J_{k-1, k}$, $\Omega_+^{(k)} = \sqrt{E_J^2 + (B_{y,1}^{(k)})^2}$, $\Omega_-^{(k)} = \sqrt{(\Delta J)^2 + (B_{y,1}^{(k)})^2}$, and the functions
\begin{align}
    f_{\pm} = \frac{1}{\Omega_{\pm}^{(k)}} \sin(\frac{\Omega_{\pm}^{(k)} t}{4}),& \hspace{1cm} 
    g_{\pm} = \cos(\frac{\Omega_{\pm}^{(k)} t}{4}),\label{Eq:fg}
\end{align}
containing the time dependence up to a global phase.
We restricted ourselves to $\phi^{(k)}=0$. For arbitrary $\phi^{(k)}$ the time evolution is shown in Appendix~\ref{App:array1Q-time-evolution}. Equations~\eqref{Eq:arrayYgate-time-evolution} and \eqref{Eq:fg} describe the time evolution analogously to Eqs.~\eqref{Eq:twodrivesphi00} and \eqref{eq:fgdef} in case of a single drive and a second neighbor. However, in contrast to Eq.~\eqref{eq:fgdef}, here $f_{\pm}$ and $g_{\pm}$ as introduced in Eq.~\eqref{Eq:fg} are defined by the difference and sum of the nearest neighbor exchange instead of the driving amplitude. 
\begin{figure*}
    \centering
    \includegraphics[width=0.98\textwidth]{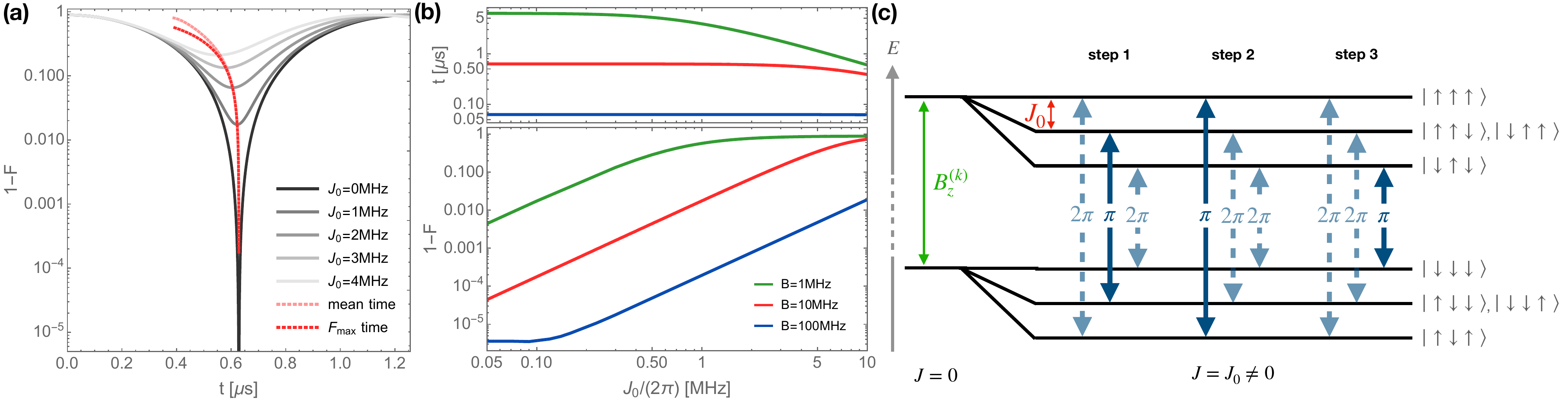}
    \caption{(a) Numerical time-dependent infidelity $1-F$ for an $IYI$ gate when driving qubit $k$ with $B_{y,1}^{(k)} = 10$ MHz. The minimum and thus optimal driving time shifts for increasing $J_0$. The light red dashed line corresponds to the mean time $\tau_Y = \pi/B_{y,1}^{(k)} + \pi/\sqrt{(B_{y,1}^{(k)})^2+J_0^2}$ and the time $F_{\rm max}$ corresponds to the numerical maximum of Eq.~\eqref{Eq:IYIfid}. (b) Optimal driving times (upper plot) and infidelity (lower plot) depending on the exchange value for $B_{y,1}^{(k)} = 1, 10, 100$~MHz. Here we chose $B_{z,1} = 20$~GHz, $B_{z,2} = 20.2$~GHz and $B_{z,3} = 20.4$~GHz. (c) $IYI$ gate in 3 steps: Each transition frequency (solid arrows) is addressed separately while synchronizing the off-resonant transitions (dashed arrows). The energy of the depicted states are split due to magnetic field $B_z^{(k)}$ and residual exchange coupling~$J_0$.}
    \label{fig:IYI}
\end{figure*}
The fidelity of performing a $Y$ gate on qubit $k$ is given by 
$F_{Y^{(k)}} = (d + |\text{Tr}[U_{i,j\neq k} (t) U_k(t)Y^{(k)}]|^2)/(d (d+1))$ with $d=2^N$. Since $U_{i,j\neq k}$ is purely diagonal we can even find an upper bound $F_{Y^{(k)}} \leq (d + |\text{Tr}_{\text{abs}}[ U_k(t)Y^{(k)}]|^2)/(d (d+1))$,
where $\text{Tr}_{\text{abs}}$ is defined in Appendix~\ref{App:fidelitybound}. Here we emphasize that for a given driving time $\tau$ we can always choose $J_{ij} = 4 m \pi/\tau$ with integer $m$, such that $U_{i,j\neq k} (\tau)=I$. In this case we find
\begin{widetext}
    \begin{align}
        F_{Y^{(k)}}= \frac{d +|\text{Tr}[U_k(t)Y^{(k)}]|^2}{d (d+1)} = \frac{1 + \frac{d}{4} \left| \frac{B_{y,1}^{(k)}}{\Omega_+^{(k)}} \sin(\frac{\Omega_+^{(k)}}{4}t) + \frac{B_{y,1}^{(k)}}{\Omega_-^{(k)}} \sin(\frac{\Omega_-^{(k)}}{4}t) \right|^2 }{d+1} .  \label{Eq:IYIfid}
    \end{align}
\end{widetext}
In the case of, e.g., 3 spin qubits, the dimension is $d=8$. We will assume $U_{i,j\neq k} (\tau)=I$ for fidelity calculations in the remainder of this work. We find an increased fidelity for $B_{y,1}^{(k)} \gg J_{k-1,k}, J_{k,k+1}$ and driving time $\tau \approx 2 \pi/ \Omega_{\pm}^{k}$. In Fig.~\ref{fig:IYI}(a) the numerical time dependent infidelity is shown for different exchange values $J_{k-1,k} = J_{k,k+1} = J_0$. If $J_0$ is small, a sufficiently good choice for the driving time is $\tau_Y = \pi/B_{y,1}^{(k)} + \pi/\sqrt{(B_{y,1}^{(k)})^2+J_0^2}$, shown as the light red line labeled by "mean time" in Fig.~\ref{fig:IYI}(a). The exact optimal timing (red dashed line labeled with $F_{max}$ time) can be obtained by solving for the maximum of Eq.~\eqref{Eq:IYIfid} numerically. Using this numerically optimized gate time, Fig.~\ref{fig:IYI}(b) shows the $Y$ gate fidelity depending on $J_0$ for driving amplitudes $B_{y,1}^{(k)} = 1$, $10$ and $100$ MHz. Similar as in Fig.~\ref{fig:DQDfid} the corrections to the RWA flattens the fidelity of $B_{y,1}^{(k)} = 100$ MHz at low residual exchange $J_0 <0.1$ MHz. The slope however is well described by the fidelity in the RWA in Eq.~\eqref{Eq:IYIfid}.

In order to characterize the error of a single-qubit gate we calculate the error matrix $M_{\rm err}=U_k(\tau_Y) Y^{(k)}$ in Appendix~\ref{App:error-matrix-coeffs}. The error matrix $M_{\rm err}$ describes the error that is left after the operation which  can then be taken into account for following operations.

Moreover, in case of a spin qubit chain, the energy levels are split by the exchange interactions $J_{k-1,k}$ and $J_{k,k+1}$ in addition to the Zeeman splitting $B_z^{(k)}$ \cite{Russ2018} (see Fig.~\ref{fig:IYI}(c)). In particular, in the case of a $Y$ gate on qubit $k$ there exist several frequencies $B_z^{(k)}\pm J_{k-1,k} \pm J_{k,k+1}$, between conditional $Y$ rotations, that have to be addressed simultaneously. As depicted in Fig.~\ref{fig:IYI}(c) in our example those are $\omega(\ket{\uparrow \uparrow \downarrow } \leftrightarrow \ket{\uparrow \downarrow \downarrow }) = \omega(\ket{\downarrow \uparrow \uparrow } \leftrightarrow \ket{\downarrow \downarrow \uparrow })= B_z^{(k)}$, $\omega(\ket{\uparrow \uparrow \uparrow } \leftrightarrow \ket{\uparrow \downarrow \uparrow }) = B_z^{(k)} + 2 J_0$, and $\omega(\ket{\downarrow \uparrow \downarrow } \leftrightarrow \ket{\downarrow \downarrow \downarrow }) = B_z^{(k)} - 2 J_0$. So far, however, we only assumed a single driving frequency $B_z^{(k)}$. Another idea to mitigate this effect could be to drive all frequencies at the same time by choosing an appropriate drive. However, when doing so, the off-resonant transitions will not be completely mitigated. In fact, numerical simulations show that a single frequency performs better than driving with all transition frequencies at the same time. Instead of a single-shot $Y$ gate on qubit $k$, one mitigation scheme, similar as proposed in Refs.~\cite{Russ2018, Toffoli2019, Kanaar_2022} is to drive each transition separately while compensating the off-resonant rotations by synchronization, see Fig.~\ref{fig:IYI}(c). In particular, when driving with frequency $\omega^{(k)} =B_z^{(k)}$ for time $\tau =4 \pi \left(n+1/2\right)/B_{y,1}^{(k)}$ this leads to the condition $B_{y,1}^{(k)} = 2\left(2n+1\right) J_0/\sqrt{4 m^2-\left(2n+1\right)^2}$ with integers $n$ and $m$. A realistic example is then $B_{y,1}^{(k)} = 2 J_0/\sqrt{3}$ with the shortest driving time $\tau =\sqrt{3} \pi /J_0$.
However, when driving with $\omega^{(k)} = B_z^{(k)} \pm 2 J_0$ the additional condition $B_{y,1}^{(k)} = 4\left(2n+1\right) J_0/\sqrt{4 l^2-\left(2n+1\right)^2}$ with integer $l$ has no exact solution. Instead, one can approximate the second condition e.g. using $n=2$, $m=5$, $l\approx 9$, and thus $B_{y,1}^{(k)}= 2 J_0/\sqrt{3} \approx 20 J_0/\sqrt{299}$ leading to longer driving times $\tau=5 \sqrt{3}\pi/J_0$. The overall 3-step $Y$ gate then has a driving time of $\tau_{Y} = 11 \sqrt{3} \pi /J_0$. Using this set of parameters we estimate a fidelity $F = (1+ d \sin(5 \sqrt{13} \pi/4)^4)/(d+1) \approx (1+ 0.99905 d)/(d+1)$. In the case of $N=3$ qubits it yields $F \approx 0.99916$. 
We can then assume charge noise on the exchange value $J_0$ as Gaussian distributed quasi static noise with standard deviation $\sigma_{J}$ and roughly estimate a fidelity decay $\propto \exp(-\sigma_J^2 \tau_Y^2/2) $. Due to the increased driving time this mitigation scheme can improve the $Y$ gate fidelity only for sufficiently low charge and dephasing noise. 
We further note that for increased exchange interaction $J_0$ additionally super exchange between qubit $k-1$ and $k+1$ lowers the energy of the $\ket{\uparrow \uparrow \downarrow }$, $\ket{\downarrow \uparrow \uparrow }$, $\ket{\uparrow \downarrow \downarrow }$ and $\ket{\downarrow \downarrow \uparrow }$ states in Fig.~\ref{fig:IYI}(c). The energy shift is equal for all these states, and thus, does not affect the transition frequencies.

\subsubsection{Extension to two-dimensional arrays}
In order to estimate the effect of residual exchange in 2D arrays we calculate the $Y$ gate fidelity analogously to Eq.~\eqref{Eq:IYIfid} when qubit $k$ has a total of four exchange-coupled neighbors, as shown in Fig.~\ref{fig:qubit_array}(b). Assuming the same value $J_0$ for all exchange couplings the fidelity again has the form $F_{Y^{(k)}} = (d+ |\text{Tr}[U_{i,j\neq k} (t) U_k(t)Y^{(k)}]|^2)/(d(d+1))$ and with $U_{i,j\neq k} (\tau)=I$, as before, simplifies to
\begin{widetext}
    \begin{align}
        F_{Y^{(k)}}=\frac{d+|\text{Tr}[U_k(t)Y^{(k)}]|^2}{d (d+1)} 
        = \frac{1+ \frac{d}{2^6} \left| 3 \sin(\frac{B_{y,1}^{(k)} t}{4}) +  \frac{4 B_{y,1}^{(k)}}{\Omega_4} \sin(\frac{\Omega_4}{4}t) + \frac{B_{y,1}^{(k)}}{\tilde{\Omega}_4} \sin(\frac{\tilde{\Omega}_4}{4}t) \right|^2}{d+1},
    \end{align}
\end{widetext}
with $\Omega_4=\sqrt{(B_{y,1}^{(k)})^2 + 4J_0^2}$ and $\tilde{\Omega}_4=\sqrt{(B_{y,1}^{(k)})^2 + 16 J_0^2}$. In contrast to the case with two neighbors, we find additional off-resonant transitions and thus obtain Rabi frequencies $\Omega_4$ and $\tilde{\Omega}_4$. A synchronization of off-resonant drives in several driving steps as explained for the case with two neighbors would require additional steps and even longer gate times than in the shown case. Increasing qubit fidelities, hence, remains a hardware problem with the aim of generating pulse amplitudes much stronger than the residual exchange between two qubits or advanced pulse shaping techniques~\cite{Kanaar_2022} to stay beneath a certain error threshold. 

\subsubsection{Drive on all qubits simultaneously}
\begin{figure}[ht]
    \centering
    \includegraphics[width=0.48\textwidth]{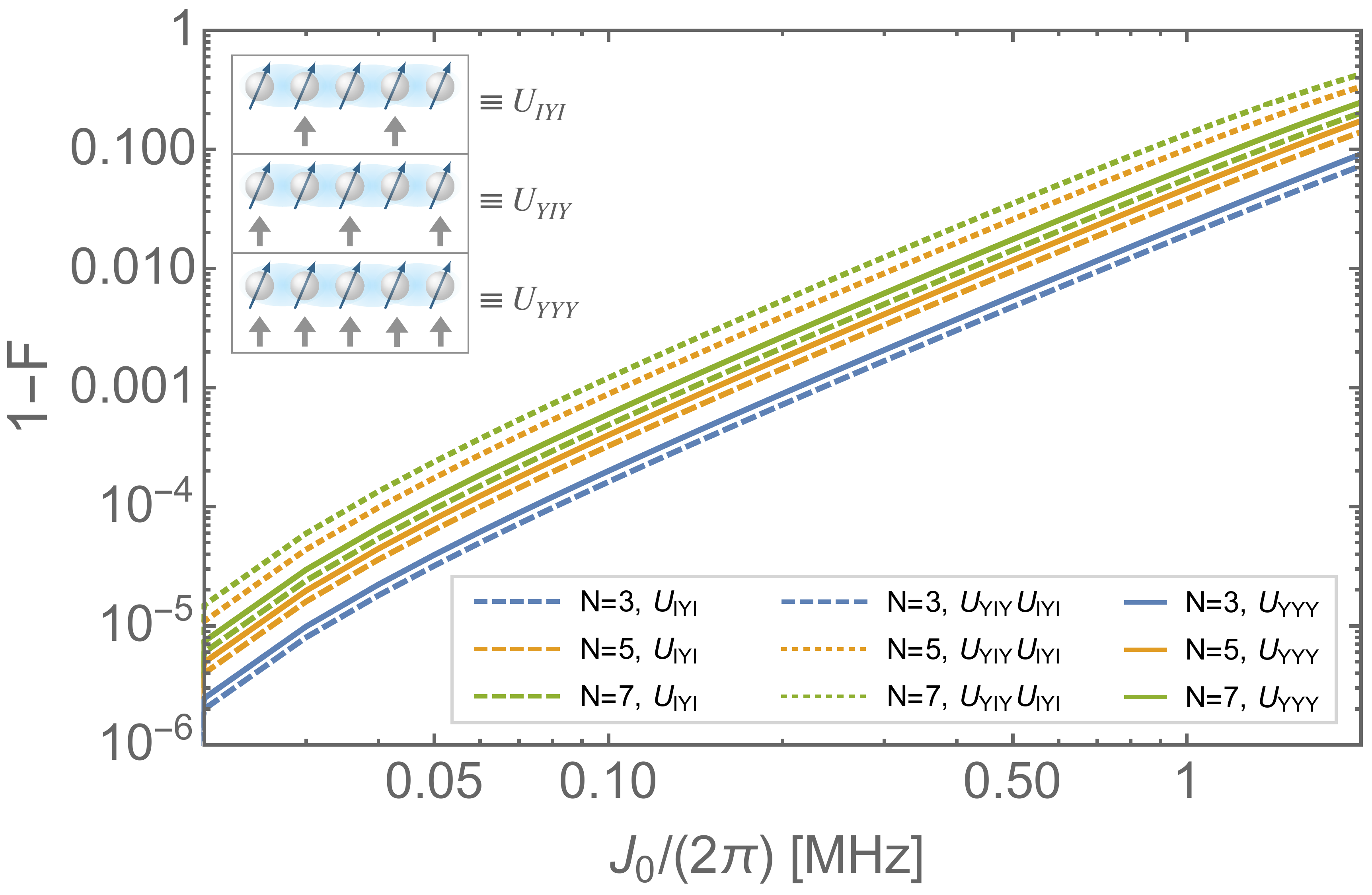}
    \caption{Residual-exchange dependent fidelity for a $Y$ gate on every second inner qubit ($U_{IYI}$), a $Y$ gate on all qubits in two steps ($U_{YIY}U_{IYI}$) and a simultaneous drive on all qubits ($U_{YYY}$). Different colors indicate the number of qubits $N=3,5,7$. In case of $N=3$ qubits the fidelity of $U_{IYI}$ and $U_{YIY}U_{IYI}$ are equal, since we use the approximated Hamiltonian and the outer $Y$ gates can be realised as described in the in the DQD case. The driving amplitude was chosen as $B_{y,1}^{(i)}=10$~MHz for all qubits $i=1, ..., N$ and the fidelity scaled to a Hilbert space of dimension $d=2^{10}$- space for comparison. The fidelity is calculated using the Hamiltonian obtained in the RWA. The inset shows which qubit drives correspond to the respective time evolution for $N=5$.}
    \label{fig:YYY-3-5-7}
\end{figure}

We have shown that at the cost of larger driving times it is in principle possible to control spin qubit arrays with residual exchange interactions. However, the algorithms that are executable on a quantum processor are limited by the relevant gate times and the qubit lifetimes in the device. Optimal control parameters such as a large driving strength increase the $Y$ gate fidelity. Similar to the case of the DQD, we now consider a simultaneous drive on several qubits. Since an analytical treatment of a system with $N \geq 3$ qubits results in a non-trivial solution for the eigenvalues and eigenvectors, we numerically calculate the approximate fidelity of a simultaneously driven $Y$ gate on 3, 5 and 7 spin qubits in Fig.~\ref{fig:YYY-3-5-7} as blue, yellow, and green solid lines, respectively (labeled as $U_{YYY}$). Here we use a driving amplitude $B_{y,1}^{(i)}=10$~MHz for all qubits $i=1, ..., N$ and optimize the driving time by maximizing the fidelity numerically. As an example, for $N=7$ and $J_0=1$~MHz the optimal time is found to be $\tau=0.624$ $\mu$s. 
We further estimate the simultaneous $Y$ gate fidelity on all qubits and compare it to the performance of driving every second inner qubit in Fig.~\ref{fig:YYY-3-5-7}, i.e., qubit 2 for $N=3$, qubits 2 and 4 for $N=5$ (see upper inset of Fig.~\ref{fig:YYY-3-5-7}), and qubits 2, 4 and 6 for $N=7$, simultaneously with numerically optimized gate times using Eq.~\eqref{Eq:IYIfid}. For this calculation the Hamiltonian obtained in the RWA was used as we found that the RWA is sufficient for a driving field of $B_{y,1}^{(i)}=10$~MHz. Thus, in the shown infidelites only slight corrections to the RWA in the regime below $J<0.05$ are expected. The fidelity is shown as dashed lines (labeled as $U_{IYI}$) lying slightly below the $U_{YYY}$ values. We also show the fidelity of a $Y$ gate on all qubits when first driving every second inner spin ($U_{IYI}$) and then all other spins ($U_{YIY}$ as shown in the inset for $N=5$) in dotted lines labeled as $U_{YIY}U_{IYI}$. In this case we assume the outer $Y$ gates to be perfect, since these can be mitigated by the $ZX$ gate as discussed for the DQD in Sec.~\ref{SubSec:DQD-mitigate}. We find that the fidelity for $U_{IYI}$ is slightly larger than for the $U_{YYY}$ case, and both fidelities are larger than the fidelity of the $U_{YIY}U_{IYI}$ case. We conclude that in residually exchange-coupled spin qubit arrays a gate on all qubits is better performed by directly driving all qubits instead of performing multiple steps because in the absence of compensation between two gate steps, errors can spread and affect the fidelity of future operations. 

\subsection{SWAP gate in qubit arrays} \label{Subsec:Array:SWAP}
So far we have found that, generally, CPHASE gates are ideally possible in an array with residual exchange. Since we assume nearest neighbor connectivity in a linear chain, the SWAP gate enables entanglement between any pairs of qubits in the array. Here we consider the effect of residual exchange on a single-pulse SWAP gate. To realize such a native SWAP gate on qubit pair $k$ and $k+1$ (where for generality we assume $1<k<N-1$), the exchange interaction $J_{k, k+1}$ is intentionally switched on, while the driving term is switched off ($B_{y,1}^{(i)} = 0$ for all~$i$). Ideally, a $\pi$-pulse on the exchange interaction between the qubits then corresponds to swapping the respective qubit states. In contrast to the previously described gates, the operation regime for such a SWAP gate requires $J\gg \Delta B_z$. The gray box in Fig.~\ref{fig:qubit_array}(c) shows such an intentional exchange pulse between qubits $k+3$ and $k+4$ (instead of $k$ and $k+1$) in a residually coupled qubit array.
In the rotating frame $ \tilde{H}(t) = R^{\dagger}HR+i\dot{R}^{\dagger}R$ with $R = \exp\Bigl(-it \sum_{i\neq k,k+1}^N B_{z}^{(i)} S^{(i)}_{z} -it (B_{z}^{(k)}+B_{z}^{(k+1)}) (S^{(k)}_{z}+S^{(k+1)}_{z})/2 \Bigr)$ we can find a time-independent Hamiltonian in the RWA $\tilde{H} \approx \sum_{\langle {i,j} \rangle \neq \langle k,k+1 \rangle} J_{ij} (S_z^{(i)} S_z^{(j)} - 1/4) + J_{k, k+1} (\mathbf{S}^{(k)} \cdot \mathbf{S}^{(k+1)} - 1/4) $ with $|J_{ij}| \ll 2 |B^{(i)}_{z} - B^{(j)}_{z}|$ for $i,j\neq k,k+1$ leading to the time evolution $U(t) = U_{i,j\notin \{k,k+1\}} (t) U_{k,k+1} (t) $, where
\begin{widetext}
    \begin{align}
        U_{i,j\notin \{k,k+1\}} (t) = \prod_{\langle i,j \rangle,  i,j\notin \{k,k+1\}} \exp(i\frac{J_{ij}}{4} t) \left( \cos(\frac{J_{ij}}{4} t) I - 4 i \sin(\frac{J_{ij}}{4} t) S_z^{(i)} S_z^{(j)} \right).
    \end{align}
\end{widetext}
Note the similarity of this expression to Eq.~\eqref{Eq:Uijnoteqk}. When choosing $J_{ij} = 4 l \pi /\tau$, where $l \in \mathbb{Z}$ and $\tau$ is the gate time, this results in the idle operation, thus we again assume $U_{i,j\neq k} (\tau)=I$. The operator $U_{k,k+1} (t)$ describes the time evolution between qubits $k-1$, $k$, $k+1$ and $k+2$. The residual exchange between $k-1$ and $k$, and $k+1$ and $k+2$ results in a decrease of the fidelity $F_{\rm SWAP^{k, k+1}} = (d + |\text{Tr}[U_k(t){\rm SWAP}^{(k, k+1)}]|^2)/(d (d+1))$ determined by 
\begin{widetext}
    \begin{equation}
    \begin{split}
    F_{\rm SWAP^{k, k+1}}=& \frac{1}{d+1}+ \frac{ d}{2^4(d+1)} \left( 4
    \cos^2\left(\frac{J_{k-1,k} t}{4}\right) \cos^2\left(\frac{J_{k+1,k+2} t}{4} \right) + J_{k,k+1}^2 \left( \sum_{\mu=1}^4 \frac{\sin(\frac{\Omega_{\mu} t}{4})}{\Omega_{\mu}} \right)^2 \right.\\
    & \hspace{3cm} \left. + 4 J_{k,k+1} \cos(\frac{J_{k-1,k} t}{4}) \cos(\frac{J_{k+1,k+2} t}{4}) \sin(\frac{ J_{k,k+1} t}{2}) \left( \sum_{\mu=1}^4 \frac{\sin(\Omega_{\mu} t)}{\Omega_{\mu}} \right) \right),
    \end{split}
    \end{equation}
\end{widetext}
where $\Omega_{\mu} = \sqrt{4 J_{k,k+1}^2 + (\Delta B_z \pm J_{k-1,k} \pm J_{k+1,k+2})^2}$ and $\Delta B_z = B_z^{(k+1)}-B_{z}^{(k)}$. The index $\mu \in \{1,2,3,4\}$ refers to all 4 possible combinations of $+$ and $-$ in the expression for $\Omega_{\mu}$.
For $2J_{k,k+1}\tau =\pi (n+1/2)$ with even integer $n$ the fidelity increases. The remaining oscillations with frequencies $\Omega_{\mu}$ and $J_0$ can in principle be synchronized, however such a condition significantly increases the gate time. Instead, the fidelity can be increased if the ratios $J_{k,k+1}/J_{k-1,k}$, $J_{k,k+1}/J_{k+1,k+2}$, $J_{k,k+1}/\Delta B_z$ become as large as experimentally feasible. Note, that in case of no residual exchange $\Omega_{1} = \Omega_{2} = \Omega_{3} = \Omega_{4}$.

To estimate the effect of residual exchange compared to the effect of a finite magnetic field gradient, we consider the error matrix $M_{\rm err} = U_{k,k+1}(\tau_{\rm SWAP}) U_{\rm SWAP}^\dagger$ in Appendix~\ref{App:error-matrix-coeffs}. Ultimately, we find that the main source of coherent errors in the case of a SWAP gate is the magnetic gradient field rather than the residual exchange. Both effects are suppressed by increasing the value of $J_{k,k+1}$ between the operating qubits.

\section{Conclusions}
We investigated the effect of residual exchange on the quantum gate performance of spin qubits. For the DQD we derived an analytical description for the gate time and fidelity of single and simultaneous $Y$ gates, and suggested high-fidelity mitigation schemes.  We showed that these mitigation protocols work under realistic conditions and that they  can (to some extent) also be applied to the outer qubits of a spin qubit chain.

Furthermore, we derived the time evolution of single-qubit and two-qubit gates in linear arrays suffering from residual exchange between nearest neighbors. We find that the identity operation can be realized by aligning the residual exchange couplings for a full period of $4\pi/J$. A CPHASE or CZ gate can be realized by choosing the exchange between the control and target qubits to be half of the other exchange values in the chain. On the other hand, a $Y$ gate operation on a spin with two or more exchange-coupled neighbors is always suffering from residual exchange and can be reduced by a high ratio between driving and exchange value. In case of a $Y$ gate on a single qubit with two coupled neighbors we find appropriate gate time conditions minimizing the infidelity, but also suggest a 3-step driving protocol addressing the respective frequencies resonantly while compensating for the  off-resonant transitions via synchronization and briefly discuss the case of a two-dimensional qubit array. We also provide a numerical fidelity calculation for simultaneous $Y$ gates on all spin qubits in a chain of 3, 5 and 7 quantum dots to give an estimated expectation for present and near term devices.

Finally, we studied the SWAP gate as a two qubit gate working in the regime $J \gtrsim \Delta B_z$, which breaks commutativity of exchange pairs. We provide a description for the resulting 4-qubit system and find the ratio $J/\Delta B_z$ as the dominant fidelity limiting factor compared to $J_{\rm on}/J_{\rm off}$.

In all cases discussed in this work, we showed that the performance of mitigation schemes strongly depends on relative values of magnetic driving, Zeeman field, and residual exchange, but also on crosstalk and several noise sources.  All these parameters determine the validity of the approximations made: For low crosstalk and high driving fields and Zeeman splitting compared to the residual exchange, pure driving without mitigation might be better, especially in the presence of rather high charge noise. However, when $J \approx B_{y,1}$ and charge noise is reasonable, mitigation will increase the qubit gate fidelity.
Furthermore, we suggest here to consider driving schemes with appropriate exchange interactions suppressed between qubits that are not frequently operated to further increase the fidelity of quantum algorithms in spin qubit arrays. 

This analysis assumes the Zeeman splitting to be much smaller than the energy splitting to higher orbital and valley states and thus the spin states representing the qubit to be well isolated from those. Moreover, spin-orbit interaction and anisotropic exchange play a very minor role for electrons in silicon. In other spin qubit platforms as holes in germanium this effect in non-negligible. However, if the residual exchange is small compared to the global Zeeman field, splitting the spin-up and spin-down states, transverse components of the exchange coupling are mostly suppressed and only the parallel $J_{zz}$ components contribute. Hence, the concepts discussed in this work are expected to be similarly applicable to other platforms.

Open questions that should be addressed in future work include the analysis and mitigation of the effects of residual exchange for other quantum gates such as Hadamard gates, T gates, and entangling two-qubit gates.  Furthermore, the combination of the presented mitigation schemes with optimized quantum control techniques may allow for shorter gates times and even higher gate fidelities.

Ultimately, experimental conditions will set the fidelity bounds for quantum operations. Our results provide the data needed for decision making between naive qubit gates and mitigation leading to the right optimization taking into account both coherent and incoherent errors. The ideal choice of control parameters, such that the gate fidelity stays beneath the error threshold for fault-tolerant quantum computing, will enable the implementation and up-scaling of quantum computing in spin-qubit platforms in the future.

\section*{Acknowledgments}
We would like to thank Bal\'azs Gul\'acsi and Stephen R. McMillan for the helpful discussions. 
This work has been supported by QLSI with funding from the European Union's Horizon 2020 research and innovation programme under grant agreement No.~951852 and by the Army Research Office (ARO) under grant numbers W911NF-15-1-0149 and W911NF-23-1-0104.

\appendix

\section{Double quantum dot} \label{App:DQD}
Here we provide a more detailed description of the time evolution for a DQD in the (1,1) charge regime starting from the Hamiltonian~\eqref{eq:DQDgeneralHamiltonian}. If we set $B_{y,1}^{\alpha} = B_{y,2}^{\alpha} = 0$ and move into the rotating frame $\tilde{H}(t) = \tilde{R}^{\dagger}H\tilde{R}+i\dot{\tilde{R}}^{\dagger}\tilde{R}$ with $\tilde{R}=\exp(-i t (B_{z}^L+B_{z}^R) (S^L_{z}+S^R_{z})/2 )$ we can find a time independent Hamiltonian in the RWA, leading to the time evolution
\begin{equation}\label{App:Eq:U0-2Q}
    \begin{split}
        \tilde{U}_0 (t) = &i e^{i\frac{J}{2}t} \frac{\Delta B_z}{2\Omega} \sin\left(\frac{\Omega}{2}t\right) (ZI-IZ) \\
        &+ \frac{1}{2} \left(1+e^{i\frac{J}{2}t} \cos\left(\frac{\Omega}{2}t\right)\right) II  \\
        &+ \frac{1}{2} \left(1-e^{i\frac{J}{2}t} \cos\left(\frac{\Omega}{2}t\right)\right) ZZ \\
        &-i e^{i\frac{J}{2}t} \frac{J}{2\Omega} \sin\left(\frac{\Omega}{2}t\right) (XX+YY), 
    \end{split}
\end{equation}
where $\Omega = \sqrt{\Delta B_z^2 + J^2}$. After time $\tau_{CP} = 2\pi n/ \Omega$ with $n\in \mathbb{Z}$ the resulting time evolution holds
\begin{equation}
\begin{split}
    \tilde{U}_0(\tau_{CP}) = &\frac{1}{2} \left(1-e^{i\frac{J}{2}\frac{2 n \pi}{\Omega}} (- 1)^n \right) ZZ \\
    &\hspace{0.cm} + \frac{1}{2} \left(1+e^{i\frac{J}{2}\frac{2 n \pi}{\Omega}} (- 1)^n \right) II,
\end{split}
\end{equation}
and equals a CPHASE gate up to a global phase factor and single qubit rotations \cite{Russ2018}. However, since the rotating frame for single-qubit gates is usually $R=\exp(-i t (B_{z}^L S^L_{z} + +B_{z}^R S^R_{z})/2)$, and the exchange is never switched off but at a finite value, we need to take into account the rotation from $\tilde{R}$ into $R$. Usually this is done by accounting for the $z$ rotations by the phase of a subsequent drive. Nevertheless, in the presence of residual exchange this does not simply equal a $z$ rotation as discussed in Appendix \ref{App:DQD-mitigate-1Q}. For this reason, we use the rotating frame $R$ in equation \eqref{Eq:U0-2Q} in the main text and assume $J$ to be small, such that we can neglect the additional terms.

The GST results for the identity gate neglecting noise and stochastic errors can be extracted as $\ln(U_{\rm actual} U_{\rm ideal}^{-1}) = \ln(U_{\rm actual}) = -i\tilde{H}t
= -i t (w_{XX} XX + w_{YY} YY + w_{ZZ} ZZ + d II + c_1 ZI + c_2 IZ)$, where $w_{XX}$, $w_{YY}$, $w_{ZZ}$, $d$, $c_1$ and $c_2$ are coefficients of the respective gate and do not have to be equal to the coefficients in the original Hamiltonian since the imaginary exponential function is not bijective. However, as discussed in the main text, it is not trivial to change rotating frames between operations with always present exchange. Thus when applying pulse sequences, that have ideal gate solutions in different rotating frames, errors might occur from these frame changes due to inappropriate pulse timing.

As a next step, we find a generalized expression for the time evolution corresponding to the Hamiltonian in Eq.~\eqref{eq:doubledot-1Qgate} in the presence of driving. Allowing $\phi_1$ and $\phi_2$ to be arbitrary angles, the time evolution is given by
\begin{widetext}
    \begin{equation}
        \begin{split}
    U (t) =  e^{-i \tilde{H} t} 
    =& \frac{1}{2} e^{i\frac{J}{4}t} \left(
    +i \sin(\phi_1) \left(E_y f^+ - \Delta B_y f^- \right) XI
    +i \sin(\phi_2)\left(E_y f^+ + \Delta B_y f^- \right) IX \right. \\
    &\hspace{1.cm}\left.
    -i \cos(\phi_1) \left(E_y f^+ - \Delta B_y f^- \right) YI
    -i \cos(\phi_2)\left(E_y f^+ + \Delta B_y f^- \right) IY \right. \\
    &\hspace{1.cm}\left.
    +\left(\sin(\phi_1) \sin(\phi_2) \left(g^+ - g^- \right) + i J \cos(\phi_1) \cos(\phi_2) \left(f^+ - f^- \right) \right) XX \right. \\
    &\hspace{1.cm}\left.
    +\left(\cos(\phi_1) \cos(\phi_2) \left(g^+ - g^- \right) + i J \sin(\phi_1) \sin(\phi_2) \left(f^+ - f^- \right) \right) YY \right. \\
    &\hspace{1.cm}\left.
    +\left(\sin(\phi_1) \cos(\phi_2) \left(-g^+ + g^- \right) + i J \cos(\phi_1) \sin(\phi_2) \left(f^+ - f^- \right) \right) XY \right. \\
    &\hspace{1.cm}\left.
    +\left(\cos(\phi_1) \sin(\phi_2) \left(-g^+ + g^- \right) + i J \sin(\phi_1) \cos(\phi_2) \left(f^+ - f^- \right) \right) YX \right. \\
    &\hspace{1.cm}\left.
    - i J \cos(\phi_1) \cos(\phi_2) \left(f^+ + f^- \right)  ZZ 
    + \left(g^+ + g^- \right) II \right), \label{Eq:app:twodrivesgeneral}
    \end{split}
    \end{equation}
\end{widetext}
where $f^{\pm}$ and $g^{\pm}$ are defined as in Eq.~\eqref{eq:fgdef}. One can calculate GST error bars again by $\ln(U_{\rm actual} (IY)^{-1})$. In particular, if only driving one qubit we obtain a time evolution given by 
\begin{align}
    U_2 (t) = i e^{i\frac{J}{4}t} \left( -B_{y,2}^R f IY - J f ZZ + g II \right) 
    \label{Eq:app:DQD-1drive}
\end{align}
with $f=f^+|_{B_{y,1}^{\alpha} = 0} = f^-|_{B_{y,1}^{\alpha} = 0}$ and $g=g^+|_{B_{y,1}^{\alpha} = 0} = g^-|_{B_{y,1}^{\alpha} = 0}$ since $\Omega_2 = \Omega_{y}^{+}|_{B_{y,1}^{\alpha} = 0} = \Omega_{y}^{-}|_{B_{y,1}^{\alpha} = 0}$. GST Hamiltonian errors are then mainly given by the coefficients of $ \ln(U_{\rm actual} (IY)^{-1}) =w_{YI}(t) YI + w_{ZX}(t) ZX + d(t) II$.

\section{Upper fidelity bound neglecting z-rotations} \label{App:fidelitybound}
We assume that the time evolution $U_{\text{actual}}$ can be calculated via a solution of the time-depedent Schroedinger equation. To quantify how good the actual time evolution matches the desired gate operation $U_{\text{ideal}}$ the fidelity is usually calculated by $F~=~(d+| \text{Tr}[ U_{\text{ideal}}^{\dagger}U_{\text{actual}} ] |^2 )/(d(d+1))$, where $d$ is the dimension of the Hilbert space. One might say it gives the amount $c_{0,\hdots,0}$ of identity of the matrix $M=U_{\text{actual}} U_{\text{ideal}}^{\dagger} = \sum_{i_1,\hdots,i_N = 0}^3 c_{i_1,\hdots,i_N} \bigotimes_n^N \sigma_{i_n}$, where $\sigma_{i_n}$ are the Pauli matrices of qubit $n$ with $i_n \in \{ 0,1,2,3\}$ and $N$ is the number of qubits. Since the Pauli matrices $\sigma_1$, $\sigma_2$ and $\sigma_3$ are traceless and $\text{Tr}[M_1\otimes M_2] = \text{Tr}[M_1] \text{Tr}[M_2]$ for any matrices $M_1$ and $M_2$. 
Nevertheless, if we assume $z$-rotations to be trivial and easily implemented, it is advantageous to find an invariant quantity under $z$-rotations. These are in general given by $R_z(\vec{\theta}) = \bigotimes_{n=1}^N \exp(i \theta_i \sigma_{3}/2)$ where $N$ is the number of qubits and $\vec{\theta} = (\theta_1,\hdots ,\theta_N)$ are the rotation angles. Note that $z$-rotations are always purely diagonal. But do not form the complete diagonal subspace of the vector space of $2^n \times 2^n$ unitary matrices.

Let us assume that a diagonal unitary $d\times d$ matrix $D$ is multiplied to another unitary matrix $M$ with same dimensions. Then $\text{Tr}[DM] = \sum_{k=1}^{d} D_{kk} M_{kk}$ does in general not equal $\text{Tr}[M]$. However, since $D$ is a diagonal unitary matrix, one has $|D_{kk}| = 1$ and thus
\begin{align}
    \text{Tr}_{\text{abs}}[DM] := \sum_{k}^{d} |D_{kk} M_{kk}| 
    = \sum_{k}^{d} |M_{kk}|
\end{align} 
stays invariant, i.e. $\text{Tr}_{\text{abs}}[DM] = \text{Tr}_{\text{abs}}[M]$. Obviously, $|\text{Tr}[DM]| = |\sum_{k}^{d} D_{kk} M_{kk}| \leq \sum_{k}^{d} |D_{kk} M_{kk}| = \sum_{k}^{d} |M_{kk}| = \text{Tr}_{\text{abs}}[DM]$.

Hence, let us consider a $z$-rotation $R_z(\theta)$ acting on the actual time evolution $U_{\text{actual}}$,
\begin{align}
    | \text{Tr}[ U_{\text{ideal}}^{\dagger}R_z(\theta) U_{\text{actual}}]| &= | \text{Tr}[ R_z(\theta) U_{\text{actual}} U_{\text{ideal}}^{\dagger} ]| \\
    &= | \text{Tr}[ R_z(\theta) M ]|\\
    &\leq \text{Tr}_{\text{abs}}[R_z(\theta) M] = \text{Tr}_{\text{abs}}[M].
\end{align}
If we find a $z$ rotation $R_z(\vec{\theta})$, such that $F$ becomes maximal, $\text{Tr}_{\text{abs}}[M]$ is invariant. In the case where a $z$-rotation leads to a perfect gate up to global phases, we even find $|\text{Tr}_{\text{abs}}[M]| = \text{Tr}[R_z(\theta)M]$. However, the reverse implication is not correct. If one calculates $\text{Tr}_{\text{abs}}[M] > |\text{Tr}[M]|$ one does not necessarily find $R_z(\vec{\theta})$ only containing single qubit rotations, such that $|\text{Tr}[R_z(\theta)M]| = \text{Tr}_{\text{abs}}[M]$. Instead, we find an upper bound for the fidelity,
\begin{align}
    F \leq \frac{d + \text{Tr}_{\text{abs}}[U_{\text{actual}} U_{\text{ideal}}^{\dagger}]^2}{d (d+1)}.
\end{align}

For a single-qubit gate in the double quantum dot case in Sec.~\ref{sec:DQD}, we estimate,
\begin{align}
    \text{Tr}_{\text{abs}}[U(t) (IY)] &= 2 \left| \frac {\Delta B_y } {\Omega_y^{-}} \sin (\frac { \Omega_y^{-}} {4} t) + \frac {E_y} {\Omega_y^{+}} \sin (\frac { \Omega_y^{+}} {4} t) \right|\\
    &\leq 2 \left( \left| \frac {\Delta B_y } {\Omega_y^{-}} \right| + \left|\frac {E_y} {\Omega_y^{+}}\right| \right),
\end{align}
and thus find the fidelity bound of
\begin{align}
    F_{IY} \leq F_{IY}^{\rm max} = \frac{1}{5} + \frac{1}{5} \left( \left| \frac {\Delta B_y } {\Omega_y^{-}} \right| + \left|\frac {E_y} {\Omega_y^{+}}\right| \right)^2 .
\end{align}
In particular, when restricting to a single drive, we obtain Eq.~\eqref{Eq:IYfidbound}.

\section{Mitigation schemes for single qubit gates in a double quantum dot} \label{App:DQD-mitigate-1Q}
To mitigate unwanted $ZZ$ terms in Eq.~\eqref{Eq:DQD-IY}, a sequence including perfectly implemented $z$ rotations 
and the choice of $J=B_{y,2}^R$ leads to an $IY$ gate with a global phase factor. Similarly, the description for the $YI$, $XI$ and $IX$ gates can be obtained. 
However, $IZ$ and $ZI$ rotations are obtained by incorporating the respective phase $\phi_2=\pi/2$ into the drive and thus leaves the evolution with a residual $ZZ$ term after the sequence. When doing so, we find
\begin{widetext}
    \begin{align}
        U_2(\tau_{IY})|_{\phi_2=\frac{\pi}{2}} &= -i (-1)^{n} e^{i \frac{\pi(2n+1)J}{2\Omega_2}} \left(\frac{B_{y,2}^R}{\Omega_2} IX +\frac{J}{\Omega_2} ZZ  \right)\\
        & \neq - i (-1)^n e^{i \frac{\pi(2n+1)J}{2\Omega_2}} \left(\frac{B_{y,2}^R}{\Omega_2} IX + \frac{J}{\Omega_2} ZI  \right) =(IZ)U_2(\tau_{IY}).
    \end{align}
\end{widetext}
Apparently, these virtual rotations do not correspond to actual rotations around the $z$ axis. Instead, a mitigation technique can be implemented with a second drive on the same qubit,
\begin{widetext}
    \begin{align}
        \tilde{U}_2(\tilde{\tau}_{IY}) U_2(\tau_{IY}) &= - \frac{(-1)^{n+\tilde{n}}}{\Omega_2 \tilde{\Omega}_2} e^{i \frac{\pi(2n+1)J}{2\Omega_2} + \frac{\pi(2\tilde{n}+1)\tilde{J}}{2\tilde{\Omega}_2}} \left(B_{y,2}^R IY + J ZZ  \right) \left(\tilde{B}_{y,2}^R IY + \tilde{J} ZZ  \right) \\
        &= - \frac{(-1)^{n+\tilde{n}}}{\Omega_2 \tilde{\Omega}_2} e^{i \frac{\pi(2n+1)J}{2\Omega_2} + \frac{\pi(2\tilde{n}+1)\tilde{J}}{2\tilde{\Omega}_2}} \left( \left(B_{y,2}^R \tilde{B}_{y,2}^R + J \tilde{J} \right) II + i \left(\tilde{B}_{y,2}^R J - B_{y,2}^R \tilde{J} \right) ZX \right) \label{Eq:App:IYIYgeneral}.
    \end{align}
\end{widetext}
Choosing appropriate values as described in the main text results in Eq.~\eqref{Eq:DQD-IYIYmain} and thus in a $ZX$ gate. A subsequent native $ZZ$ gate finally leads to an $IY$ gate implementation.

\section{Sensitivity to noise} \label{Sec:DQDnoise}
Since charge noise is often the factor limiting the fidelity, we address here the sensitivity of above mitigation schemes with respect to small fluctuations of the exchange coupling in a double quantum dot. Therefore, neglecting other error sources such as, e.g., dephasing and crosstalk, we can write the fidelity for each gate in first order (assuming $\cos(x\approx 0) \approx 1$) ,
\begin{align}
    F_{IY} \approx& \frac{1}{5} + \frac{4}{5} \left| \frac{B_y^R}{\sqrt{(B_y^R)^2+J^2}} \right|^2 , \\
    F_{ZX} \approx& \frac{1}{5} + \frac{4}{5} \left| \frac{2 B_y^R J}{(B_y^R)^2+J^2} \right|^2 , \\
    F_{YY} \approx& \frac{1}{5} + \frac{4}{5} \left| \frac{J}{\sqrt{E_y^2+J^2}} - \frac{J}{\sqrt{\Delta B_y^2+J^2}} \right|^2  ,
\end{align}
and find the $ZX$ and $YY$ gates to be more sensitive to exchange noise than the conventional $IY$ gates. 
\begin{figure}[hb]
    \centering
    \includegraphics[width=0.48\textwidth]{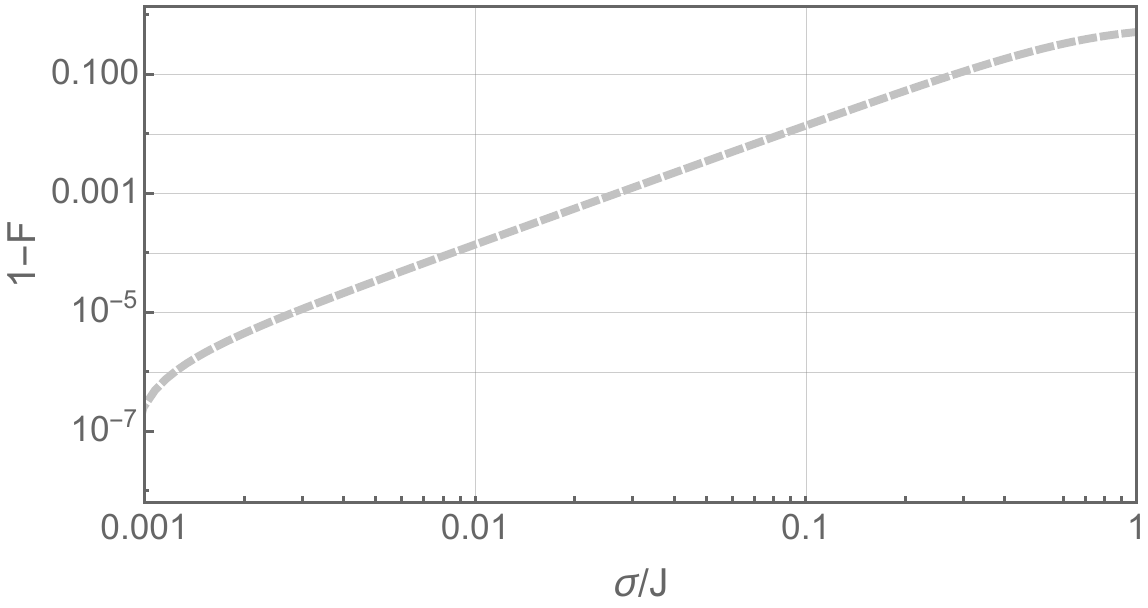}
    \caption{
    Numerical infidelity results for a DQD with residual exchange using the following parameters: $B_z^L=20$ GHz, $B_z^R=20.2$ GHz, $B_{y,0}^L=0.1$ MHz, $B_{y,0}^R=0$ MHz. }
    \label{fig:RWA-noise}
\end{figure}
Thus, our mitigation schemes only apply for reasonably low  charge noise on the exchange as shown in Fig.~\ref{fig:DQDfid}. The lower infidelity bound due to quasi-static noise on the exchange value is determined by the standard deviation to exchange ratio as shown in Fig.~\ref{fig:RWA-noise}. However, since the qubit frequency $B_z^{R}$ is also suffering from charge noise, we suggest to compare benchmarking results of both, conventional $IY$ and $YY$ gates (where $J$ should be as small as possibile) and the mitigated $ZX$ and $YY$ gates as in Eq.~\eqref{Eq:DQD-IYIYmain} and \eqref{Eq:YY} to achieve the highest possible fidelity in a specific setup.
Although the approximations made seem to be reasonable, the neglected terms can lead to a reduced infidelity when parameters sensitive to the approximations, such as $B_{y}^{\alpha}$ and $J$, are increased. Moreover, in experimental setups, longer gate times and gate sequences compete with relaxation and dephasing times of the qubit, and thus their applicability need to be checked on the respective device.

\section{Mitigation for the idle operation} \label{App:DQD-idle}
\begin{figure}[hb]
    \centering
    \includegraphics[width=0.48\textwidth]{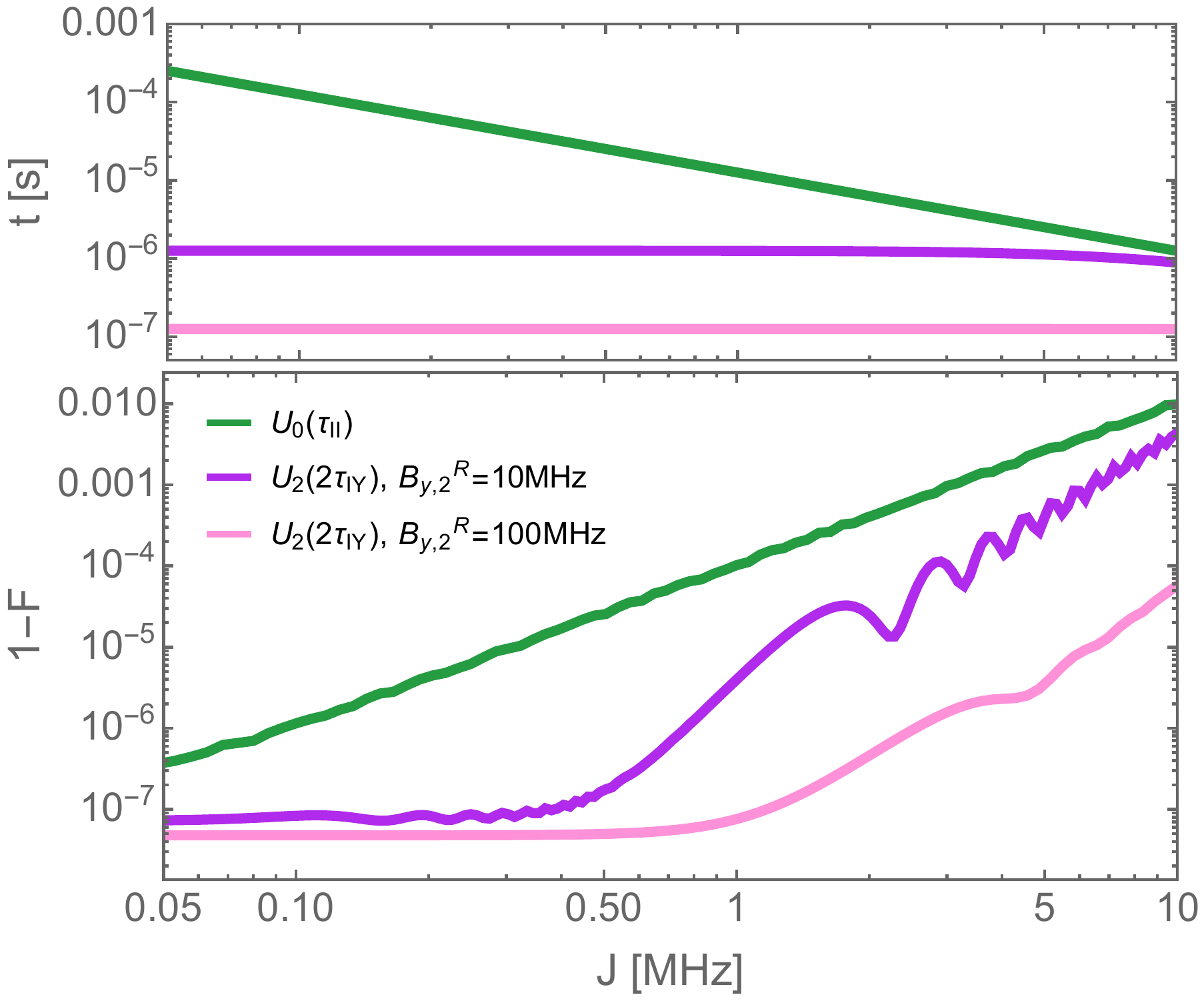}
    \caption{
    Lower plot shows numerical infidelity results for an idle gate in a DQD with residual exchange using the following parameters: $B_z^L=20$ GHz, $B_z^R=20.2$ GHz. Performance of $U_0(\tau_{II})$ in Eq.~\eqref{Eq:U0-2Q} (green) and $U_2(2\tau_{IY})$ in Eq.~\eqref{Eq:app:DQD-1drive} with two different driving amplitudes $B_{y,2}^{R}=10$ MHz and $100$ MHz are compared. The upper plots shows the corresponding driving times of the respective gates.}
    \label{fig:DQD-II-compare}
\end{figure}

In Secs.~\ref{Subsec:DQD-nodrive} and \ref{Subsec:DQD+drive} we have found two implementations of the idle operation to mitigate the effect of residual exchange. In the lower plot of Fig.~\ref{fig:DQD-II-compare} we compare the numerical gate performance of the two implementations $U_0(\tau_{II})$ in Eq.~\eqref{Eq:U0-2Q} and $U_2(2\tau_{I2})$ in Eq.~\eqref{Eq:app:DQD-1drive} for two different driving amplitudes. We find the implementation $U_2(2\tau_{IY})$ to have lower infidelity due to shorter driving times. The finite infidelity is determined by corrections to the RWA. The gate times $\tau_{II} = 4n\pi/J$ and $2\tau_{IY} = 4\pi (2n+1)/\Omega_2$ are shown in the upper plot.

\section{Time evolution for a single drive on one qubit in arrays} \label{App:array1Q-time-evolution}
In general, the time evolution of a drive on qubit $k$ with arbitrary phase $\phi^{(k)}$ is given by
\begin{widetext}
    \begin{equation}
        \begin{split}
        U_k(t) &= e^{it\frac{J_{k-1, k}+J_{k, k+1}}{4} } \exp(-it \left( J_{k-1, k} S_z^{(k-1)} S_z^{(k)}+ J_{k, k+1} S_z^{(k)} S_z^{(k+1)} + \frac{B_{y,1}}{2} \left( e^{-i \phi^{(i)}} S_{+}^{k} - e^{i \phi^{(i)}} S_{-}^{k} \right) \right))\\
        &= e^{it \frac{J_{k-1, k}+J_{k, k+1}}{4} } \frac{1}{2} \left(
        i B_{y,1}^{(k)} \left( f_+ + f_- \right) \left( \sin(\phi^{(k)}) X^{(k)}
        - \cos(\phi^{(k)}) Y^{(k)} \right) \right. \\
        &\hspace{3cm} \left.
        + i B_{y,1}^{(k)} \left( f_+ - f_- \right) \left( \sin(\phi^{(k)})  Z^{(k-1)}X^{(k)}Z^{(k+1)}
        - \cos(\phi^{(k)}) Z^{(k-1)}Y^{(k)}Z^{(k+1)} \right) \right. \\
        &\hspace{3cm} \left.
        -i \left( E_J f_+ - \Delta J f_- \right) Z^{(k-1)} Z^{(k)} 
        -i  \left( E_J f_+ + \Delta J f_- \right) Z^{(k)} Z^{(k+1)} \right. \\
        &\hspace{3cm} \left.
        +i \left( g_+ - g_- \right) Z^{(k-1)} Z^{(k+1)} 
        + \left( g_+ + g_- \right) I
        \right) ,
    \end{split}
    \end{equation}
\end{widetext}
where $X^{(i)}$, $Y^{(i)}$ and $Z^{(i)}$ are the Pauli operators on qubit~$i$. When $\phi^{(k)} = 0$ we obtain the time evolution given in Eq. \eqref{Eq:arrayYgate-time-evolution}.

\section{Error matrix coefficients} \label{App:error-matrix-coeffs}
For a better understanding of the error occurring during a non-ideal operation we calculate the error matrix $M_{\rm err} = U_{\rm actual}U^\dagger_{\rm ideal}$. It corresponds to the error that is left after the operation, i.e. $U_{\rm actual} = M_{\rm err}U_{\rm ideal}$. 

\begin{figure}[ht]
    \centering
    \includegraphics[width=0.40\textwidth]{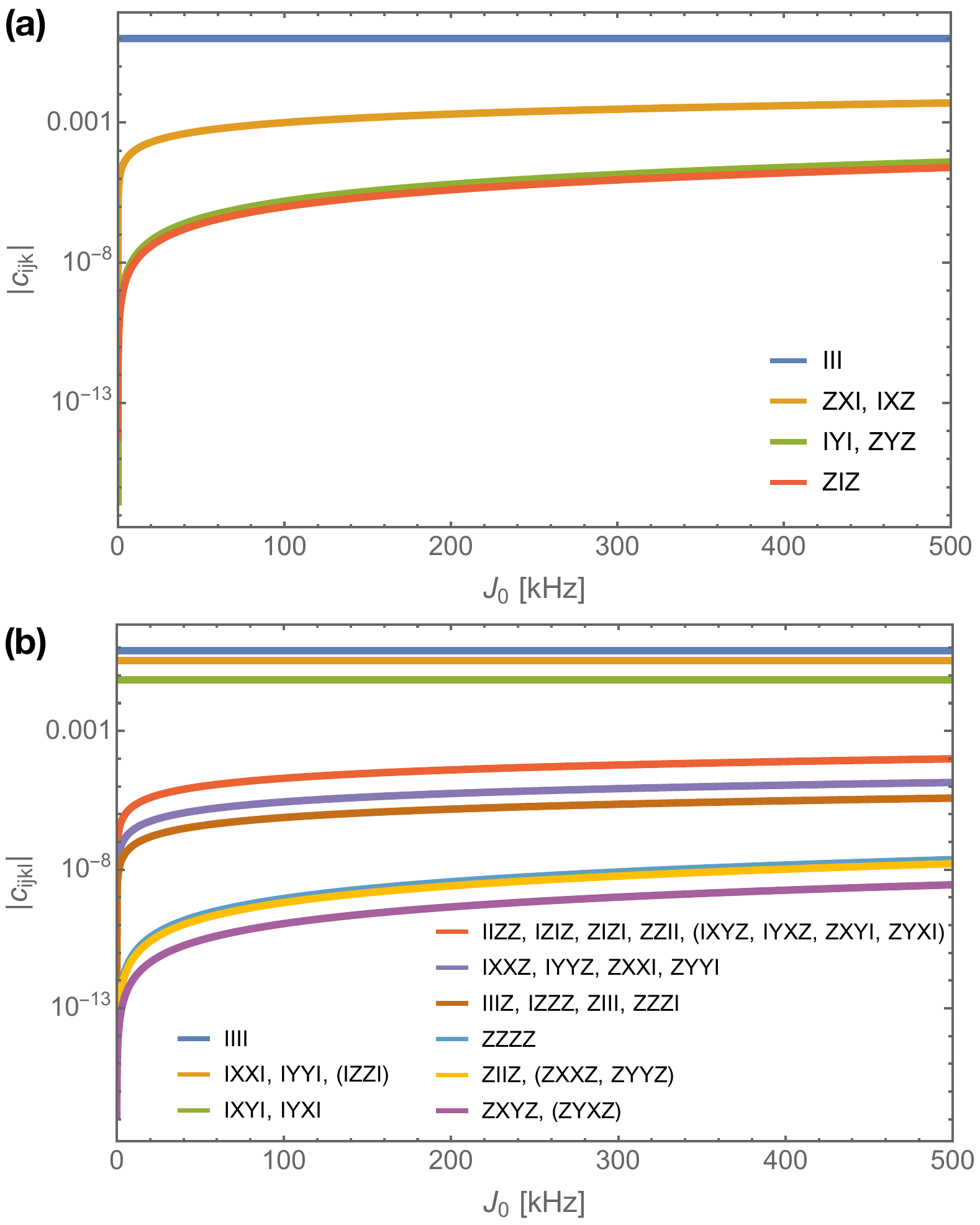}
    \caption{Absolute values of the coefficients of the error matrix $M_{\rm err}$ for (a) a single-qubit Y gate and (b) a two-qubit SWAP gate on a qubit $k$ (and $k+1$) within a qubit chain with residual exchange couplings present. We chose $B_{y,k} = 100$~MHz in (a), and $J_{k,k+1} = 1$~GHz and a gradient field of $\Delta B_z = 200$~MHz in (b).}
    \label{fig:cijkl}
\end{figure}

In case of a single $Y$ gate in a linear qubit array we can write $M_{\rm err}=U_k(\tau_Y) Y^{(k)} = \sum_{i,j,k=0}^{3} c_{ijk} \, \sigma_i \otimes \sigma_j \otimes \sigma_k$. In Fig.~\ref{fig:cijkl}(a) we plot the absolute coefficients $|c_{ijk}|$. The main error sources arise from coefficients of $ZXI$ and $IXZ$ which are generally given by 
\begin{align}
    |c_{310}| = \frac{1}{2} \left| \frac{E_J}{\Omega_+^{(k)}}  \sin(\frac{\Omega_+^{(k)}}{4}t) + \frac{\Delta J}{\Omega_-^{(k)}} \sin(\frac{\Omega_-^{(k)}}{4}t)\right|,
\end{align}
and $|c_{310}|=|c_{013}|$. To optimally control the qubit, this rotation needs to be taken into account for the following operations, or it can be reduced by increasing the driving strength $B_{y,1}^{(k)}$ and by reducing the residual exchange values $J_{k-1,k}$ and $J_{k,k+1}$.

In the case of a SWAP gate we calculate the error matrix $M_{\rm err} = U_k(\tau_{\rm SWAP}) U_{\rm SWAP}^\dagger = \sum_{i,j,k,l=0}^3 c_{ijkl} \, \sigma_i \otimes \sigma_j \otimes \sigma_k \otimes \sigma_l$ and show the contributing errors in Fig.~\ref{fig:cijkl}(b). Here, we use $J_{k-1,k}=J_{k+1,k+2}=J_0$. We find the main sources of coherent errors $IXXI$, $IYYI$, $IZZI$, $IXYI$, and $IYXI$ to be dominated by the magnetic gradient field rather than the residual exchange, which also appears in the simple two-qubit case without additional residual exchange $J_{0}=0$ to outside qubits. By increasing the intentional exchange value of $J_{kk+1}$, this effect decreases together with the effect of residual exchange. In contrast to the previously investigated single-qubit gates, a finite on-off ratio of the experimentally feasible exchange interaction does not essentially limit the performance of a native SWAP gate compared to the error due to a finite magnetic gradient.

\bibliography{bibliography}

\end{document}